\documentclass[twocolumn]{aastex631}

\usepackage{amsmath}
\usepackage{amssymb}
\usepackage{comment}

\newcommand{\M}{{\ $\mid$}}

\newcommand{\W}{{$\lambda$}}

\newcommand{\OII}{[O\,\textsc{ii}]}

\newcommand{\HI}{H\,\textsc{i}}

\newcommand{\q}{$q_{\rm ion}$}

\newcommand{\Add}[1]{\textcolor{black}{#1}}

\begin{document}
%\linenumbers
\title{
SILVERRUSH. XIV. Ly$\alpha$ Luminosity Functions and Angular Correlation Functions from\\ 
$\sim 20,000$ Ly$\alpha$ Emitters at $z\sim2.2-7.3$ \Add{from upto} 
${24 ~\rm deg}^2$ \Add{HSC-SSP and CHORUS} Surveys:\\
Linking the Post-Reionization Epoch to the Heart of Reionization
}

\author[0009-0008-0167-5129]{Hiroya Umeda}
\affiliation{Institute for Cosmic Ray Research,
The University of Tokyo,
5-1-5 Kashiwanoha, Kashiwa,
Chiba 277-8582, Japan}
\affiliation{Department of Physics, Graduate School of Science, The University of Tokyo, 7-3-1 Hongo, Bunkyo, Tokyo 113-0033, Japan}
\email{ume@icrr.u-tokyo.ac.jp}

\author[0000-0002-1049-6658]{Masami Ouchi}
\affiliation{National Astronomical Observatory of Japan, 2-21-1 Osawa, Mitaka, Tokyo 181-8588, Japan}
\affiliation{Institute for Cosmic Ray Research,
The University of Tokyo,
5-1-5 Kashiwanoha, Kashiwa,
Chiba 277-8582, Japan}
\affiliation{Department of Astronomical Science, SOKENDAI (The Graduate University for Advanced Studies), Osawa 2-21-1, Mitaka, Tokyo,
181-8588, Japan}
\affiliation{Kavli Institute for the Physics and Mathematics of the Universe (WPI), 
University of Tokyo, Kashiwa, Chiba 277-8583, Japan}

\author[0000-0003-3214-9128]{Satoshi Kikuta}
\affiliation{Department of Astronomy, School of Science, The University of Tokyo, 7-3-1 Hongo, Bunkyo, Tokyo 113-0033, Japan}

\author[0000-0002-6047-430X]{Yuichi Harikane}
\affiliation{Institute for Cosmic Ray Research,
The University of Tokyo,
5-1-5 Kashiwanoha, Kashiwa,
Chiba 277-8582, Japan}
\affiliation{Department of Physics and Astronomy, University College London, Gower Street, London WC1E 6BT, UK}

\author[0000-0001-9011-7605]{Yoshiaki Ono}
\affiliation{Institute for Cosmic Ray Research,
The University of Tokyo,
5-1-5 Kashiwanoha, Kashiwa,
Chiba 277-8582, Japan}

\author{Takatoshi Shibuya}
\affiliation{Kitami Institute of Technology, 165 Koen-cho, Kitami, Hokkaido 090-8507, Japan}

\author[0000-0002-7779-8677]{Akio K. Inoue}
\affiliation{Waseda Research Institute for Science and Engineering, Faculty of Science and Engineering, Waseda University, 3-4-1 Okubo, Shinjuku,
Tokyo 169-8555, Japan}
\affiliation{Department of Physics, School of Advanced Science and Engineering, Faculty of Science and Engineering, Waseda University, 3-4-1
Okubo, Shinjuku, Tokyo 169-8555, Japan}

\author[0000-0002-2597-2231]{Kazuhiro Shimasaku}
\affiliation{Department of Astronomy, School of Science, The University of Tokyo, 7-3-1 Hongo, Bunkyo, Tokyo 113-0033, Japan}
\affiliation{Research Center for the Early Universe, Graduate School of Science,
The University of Tokyo, 7-3-1 Hongo, Bunkyo, Tokyo 113-0033, Japan}

\author[0000-0002-2725-302X]{Yongming Liang}
\affiliation{Institute for Cosmic Ray Research,
The University of Tokyo,
5-1-5 Kashiwanoha, Kashiwa,
Chiba 277-8582, Japan}
\affiliation{National Astronomical Observatory of Japan, 2-21-1 Osawa, Mitaka, Tokyo 181-8588, Japan}

\author{Akinori Matsumoto}
\affiliation{Institute for Cosmic Ray Research,
The University of Tokyo,
5-1-5 Kashiwanoha, Kashiwa,
Chiba 277-8582, Japan}
\affiliation{Department of Physics, Graduate School of Science, The University of Tokyo, 7-3-1 Hongo, Bunkyo, Tokyo 113-0033, Japan}

\author[0000-0002-6186-5476]{Shun Saito}
\affiliation{Institute for Multi-messenger Astrophysics and Cosmology, Department of Physics, Missouri University of Science and Technology, 1315 N Pine Street, Rolla, MO 65409, U.S.A.}
\affiliation{Kavli Institute for the Physics and Mathematics of the Universe (WPI), 
University of Tokyo, Kashiwa, Chiba 277-8583, Japan}

\author[0000-0002-3801-434X]{Haruka Kusakabe}
\affiliation{National Astronomical Observatory of Japan, 2-21-1 Osawa, Mitaka, Tokyo 181-8588, Japan}

\author{Yuta Kageura}
\affiliation{Institute for Cosmic Ray Research,
The University of Tokyo,
5-1-5 Kashiwanoha, Kashiwa,
Chiba 277-8582, Japan}
\affiliation{Department of Physics, Graduate School of Science, The University of Tokyo, 7-3-1 Hongo, Bunkyo, Tokyo 113-0033, Japan}

\author[0009-0000-1999-5472]{Minami Nakane}
\affiliation{Institute for Cosmic Ray Research,
The University of Tokyo,
5-1-5 Kashiwanoha, Kashiwa,
Chiba 277-8582, Japan}
\affiliation{Department of Physics, Graduate School of Science, The University of Tokyo, 7-3-1 Hongo, Bunkyo, Tokyo 113-0033, Japan}

\begin{abstract}
We present the luminosity functions (LFs) and angular correlation functions (ACFs) derived from 18,960 Ly$\alpha$ emitters (LAEs) at $z=2.2-7.3$ over a wide survey area of $\lesssim24~{\rm deg^2}$ that are identified in the narrowband data of the Hyper Suprime-Cam Subaru Strategic Program (HSC-SSP) and the Cosmic HydrOgen Reionization Unveiled with Subaru (CHORUS) surveys. Confirming the large sample with the 241 spectroscopically identified LAEs, we determine Ly$\alpha$ LFs and ACFs in the bright\Add{er} luminosity range down to $0.5L_{\star}$, and confirm that our measurements are consistent with previous studies but offer significantly reduced statistical uncertainties. The improved precision of our ACFs allows us to clearly detect one-halo terms at some redshift\Add{s}, and provides large-scale bias measurements that indicate hosting halo masses of $\sim 10^{11} M_\odot$ over $z\simeq 2-7$. By comparing our Ly$\alpha$ LF (ACF) measurements with reionization models, we estimate the neutral hydrogen fractions in the intergalactic medium to be $x_{\rm \HI}$\Add{$<0.05$ (=${0.06}^{+0.12}_{-0.03}$), $0.15^{+0.10}_{-0.08}$ (${0.21}^{+0.19}_{-0.14}$), $0.18^{+0.14}_{-0.12}$, and $0.75^{+0.09}_{-0.13}$} at $z=5.7$, $6.6$, $7.0$, and $7.3$, respectively. Our findings suggest that the neutral hydrogen fraction remains relatively low, $x_{\rm \HI} \lesssim 0.2$, at $z=5-7$, but increases sharply at $z > 7$, reaching $x_{\rm \HI} \sim 0.9$ by $z \simeq 8-9$, as indicated by recent JWST studies. The combination of our results \Add{from LAE observations} with recent JWST observations suggests that the major epoch of reionization occurred around $z \sim 7-8$, likely driven by the emergence of \Add{massive sources emitting significant ionizing photons}.
\end{abstract}

\section{Introduction} \label{sec:intro}
Cosmic reionization is the event during which neutral hydrogen ({\HI}) in the intergalactic medium (IGM) became fully ionized in the early universe \citep[e.g.,][]{2022ARA&A..60..121R}. However, much of how cosmic reionization proceeded, what ionized the IGM, and how ionized regions spread throughout the IGM remains debated. Because the cosmic reionization history depends on the physical properties of sources driving the ionization of the IGM \citep[e.g.,][]{I18,F19,N20,Munoz24}, we need to determine its history for a comprehensive understanding of the cosmic reionization scenario. \par
One quantity often used to describe the stage of the cosmic reionization process is hydrogen neutral fraction ($x_{\rm \HI}$). The redshift evolution of the volume-averaged $x_{\rm \HI}$ is generally used to describe the cosmic reionization history. Because {\HI} gas scatters the Ly$\alpha$ line, Ly$\alpha$ transmission is a good tracer of $x_{\rm \HI}$. Optical depth measurement of Ly$\alpha$ using multiple quasar (QSO) UV spectra demonstrated that cosmic reionization has almost fully completed (i.e., $x_{\rm \HI} < 10^{-4}$) by $z\sim5$ \citep{2006AJ....132..117F}. Ly$\alpha$ transmission measurements from galaxies, QSOs, and gamma ray bursts (GRBs) provide numerous $x_{\rm \HI}$ constraints at $z=6-7$. However, because the $x_{\rm \HI}$ constraints from different methods are scattered, we do not yet have a consensus on how the cosmic reionization history proceeded.\par
Recently, high-redshift galaxy observations by the James Webb Space Telescope (JWST) have enabled the study of the epoch or reionization (EoR) at $z>7$ \citep[e.g.,][]{CL23,Br23,Hsiao23,Mo23,Umeda24,Nakane24}. JWST observations greatly extend the redshift frontier of $x_{\rm \HI}$ up to $z\sim12$. However, the selection of sightlines used to infer $x_{\rm \HI}$ at $z>7$ is most likely biased, and the number of sightlines is still limited. \cite{Ishimoto22} compare spatial distribution between high IGM transmission and ionizing sources (i.e., galaxies) using background quasar spectra and foreground galaxy. \cite{Ishimoto22} find that IGM has high Ly$\alpha$ transmission where galaxy density is high, which suggests that cosmic reionization proceeds spatially inhomogeneously. Spatial inhomogeneity in the ionization state of the IGM introduces larger variance in the Ly$\alpha$ transmission among sightlines. \cite{MF08a} find that such an inhomogeneity in IGM ionization state could introduce up to a 30\% in $x_{\rm \HI}$ systematics compared to a homogeneous IGM at the middle stage of cosmic reionization. To account for such an effect introduced by inhomogeneity in the IGM, we need Ly$\alpha$ transmission measurements from multiple unbiasedly-selected sightlines in a wide survey field. \par

One possible way is to extract Ly$\alpha$ transmission using a blind wide-field galaxy survey. There have been multiple surveys for Ly$\alpha$ emitters (LAEs) both photometrically and spectroscopically \citep[e.g.,][]{Ouchi08,Ouchi18,Inoue20,Drake17,Taylor21}. Photometric surveys for LAE often use narrow-band filters to detect strong Ly$\alpha$ emission. With the blind wide-field survey for LAEs, we can infer statistical properties of LAEs such as Ly$\alpha$ luminosity functions and clusterings \citep[e.g.,][]{Ouchi20}. During the EoR, Ly$\alpha$ luminosity functions evolve with redshift due to the effect of cosmic reionization and intrinsic galaxy evolution, similar to that observed in Lyman break galaxies (LBGs). The clustering properties of LAEs also evolve with redshift due to the non-negligible redshift evolution of the IGM optical depth around galaxies during the EoR. Many theoretical models suggest the redshift evolution of Ly$\alpha$ luminosity function and clusterings are good tracers to determine the global $x_{\rm \HI}$ values at different redshifts \citep[e.g., ][]{Furlanetto06,McQuinn07,Sobacchi15,2018PASJ...70...55I,Garel21,Morales21,SalvadorSole22,Xu23}. \par

In this work, we present new measurements of the Ly$\alpha$ luminosity functions at $z\sim2.2-7.3$ and clustering at $z\sim2.2-6.6$ based on deep and ultra-deep optical images obtained in the Hyper Suprime-Cam (HSC) Subaru Strategic Program \citep[SSP;][]{Aihara18a,Aihara18b,Aihara19,Aihara22} survey and the Cosmic HydrOgen Reionization Unveiled with Subaru \citep[CHORUS;][]{Inoue20}. This paper is a part of Systemic Identification of LAEs for Visible Exploration and Reionization Research Using Subaru HSC \cite[SILVERRUSH;][]{Ouchi18,Shibuya18a,Shibuya18b,Konno18,Harikane18,2018PASJ...70...55I,Higuchi19,Harikane19,Kakuma21,Ono21,Goto21,Kikuchihara22,Kikuta23}. \cite{Kikuta23} have constructed LAE samples at $z=2.2-7.3$ based on the photometrically identified LAE candidates via the SILVERRUSH program. We use the LAE catalogs of Kikuta et al to calculate the Ly$\alpha$ luminosity function and two-point auto-correlation function (ACF) at each redshift. We then infer $x_{\rm \HI}$ at $z\gtrsim6$ from the redshift evolution of the Ly$\alpha$ luminosity function and ACF by comparing our results with the predictions from simulations.\par 

This paper is organized as follows. We explain the observational data sets and how we select our samples in Section \ref{sec:sample}. We present our results of the Ly$\alpha$ luminosity functions in Section \ref{sec:lf} and the clustering analysis in Section \ref{sec:cluster}. We discuss our findings, including the application to constraining cosmic reionization history in Section \ref{sec:disc}. We summarize our findings in Section \ref{sec:conc}. In this paper, we adopt a cosmological parameter set from TT + TE + EE + lowE + BAO + lensing results in Table 2 of \cite{Planck20}: $h=0.6766$, $\Omega_m=0.3103$, $\Omega_\Lambda=0.6897$, $\Omega_b h^2=0.02234$, and $Y_p=0.248$. All magnitudes are in the AB system. 

\section{Observations and sample selection } \label{sec:sample}

\subsection{Hyper Suprime-Cam imaging observations and data reduction}
\begin{figure}[htbp]
    \includegraphics[width=\linewidth]{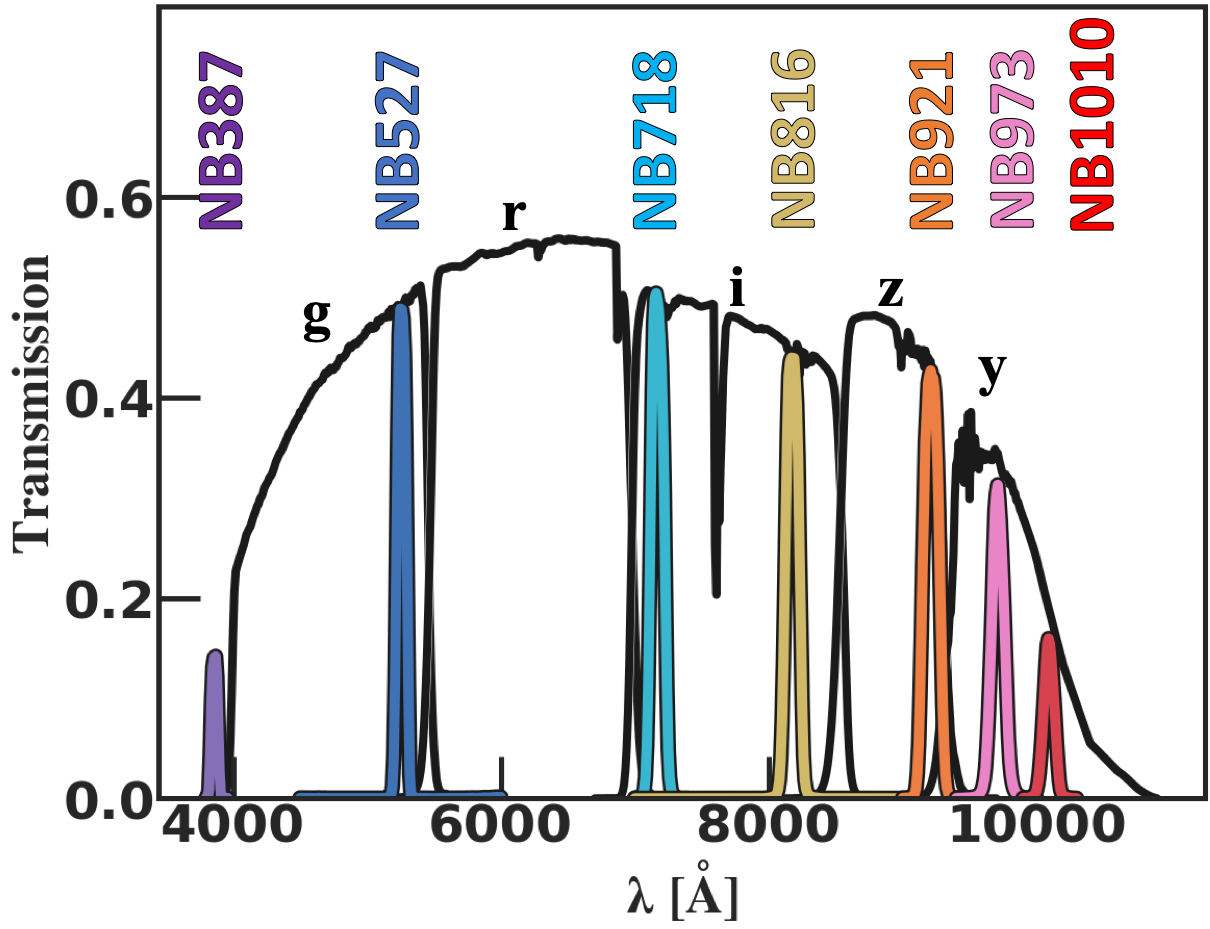}
    \label{transmission_curve}
    \caption{Transmission curves of Subaru/HSC broad and narrow band filters. The black lines represent the curves for the broad band filter ($g$, $r2$, $i2$, $z$, and $y$). Note that the $r$- and $i$-band filters are replaced with the new $r2$ and $i2$ filters in 2016. For simplicity, we refer to them as $r$ and $i$ bands. The purple, blue, pale-blue, yellow, orange, pink, and red lines represent the curves for the $NB387$, $NB527$, $NB718$, $NB816$, $NB921$, $NB973$, and $NB1010$ narrow band filters, respectively. }
\end{figure}

We construct LAE samples based on the photometric catalog presented in \cite{Kikuta23}. \cite{Kikuta23} construct LAE samples based on Subaru/HSC narrowband and broadband imaging data from HSC-SSP S21A internal release and the data from the CHORUS project. HSC-SSP imaging data consists of Wide, Deep (D), and UltraDeep (UD) layers. Narrow-band imaging data for $NB387$, $NB816$, and $NB921$ filters have been obtained for the Deep layer. $NB816$, $NB921$, and $NB1010$ filter imaging data are taken in the UD layer. We refer to the fields covered by the Deep (UD) layer in COSMOS, DEEP23, ELAISN1 and SXDS as D\Add{-}COSMOS (UD\Add{-}COSMOS), \Add{D-}DEEP23, \Add{D-}ELAISN1, and D\Add{-}SXDS (UD\Add{-}SXDS), respectively. 
In this work, we strictly define the UD-COSMOS and UD-SXDS fields in the $NB816$ and $NB921$ imaging data. UD-COSMOS is defined as the area within a circle centered at (10:00:28.6008, +2:12:20.986) with a radius of 3100 arcsec, and UD-SXDS is defined similarly with a center at (2:17:58.9008, -4:52:23.50). The imaging data from the CHORUS project consist of the deep image that covers almost the same area as the UD\Add{-}COSMOS field in HSC-SSP. Therefore, we simply refer to the survey field of \Add{the} CHORUS project as UD\Add{-}COSMOS. Narrow-band imaging data of $NB387$, $NB527$, $NB718$, and $NB973$ filters have been taken in the CHORUS project. All of the HSC-SSP and CHORUS fields are covered by imaging data from the $g$, $r$, $i$, $z$, and $y$ broadband filters. The redshift of Ly$\alpha$ \Add{emission} at the central wavelength of $NB387$, $NB527$, $NB718$, $NB816$, $NB921$, $NB973$, and $NB1010$ corresponds to $z=2.2$, 3.3, 4.9, 5.7, 6.6, 7.0, and 7.3, respectively. Hereafter, we refer to LAE samples selected based on $NB387$, $NB527$, $NB718$, $NB816$, $NB921$, $NB973$, and $NB1010$ imaging data as $z=2.2$, 3.3, 4.9, 5.7, 6.6, 7.0, and 7.3 LAE samples, respectively. We summarize the size of the survey area in each field and transmission curves for the broad and narrow filters in Table \ref{tab:surveyarea} and Figure \ref{transmission_curve}, respectively. Please refer to \cite{Kikuta23} for more details on the properties and reduction process of the HSC-SSP and CHORUS imaging data.\par
After the reduction of imaging data described in \cite{Kikuta23}, we apply the sample selection using the narrowband magnitude and narrowband color excess due to Ly$\alpha$ emission line. We adopt the \Add{same} selection criteria as those adopted by \cite{Kikuta23} \Add{except for the $z=6.6$ LAE sample}:

\begin{eqnarray}
\begin{split}
\mathrm{NB387} < \mathrm{NB387_{6\sigma}}~\mathrm{AND} \\
\mathrm{g} - \mathrm{NB387} > \mathrm{max}(0.2, (\mathrm{g}-\mathrm{NB387})_\mathrm{3\sigma})~\mathrm{AND} \\
\mathrm{r} - \mathrm{NB387} > 0.3~\mathrm{AND} \\
\mathrm{i} - \mathrm{NB387} > 0.4~\mathrm{AND} \\
\mathrm{z} - \mathrm{NB387} > 0.5
\end{split}
\end{eqnarray}
for $z=2.2$ LAEs, 
\begin{eqnarray}
\begin{split}
\mathrm{NB527}  <  \mathrm{NB527_{5\sigma}}~\mathrm{AND} \\
\mathrm{g} - \mathrm{NB527} > \mathrm{max}(0.9, (\mathrm{g}-\mathrm{NB527})_\mathrm{3\sigma})~\mathrm{AND} \\
\mathrm{r} - \mathrm{NB527} > 0.3~\mathrm{AND} \\
\mathrm{i} - \mathrm{NB527} > 0.4~\mathrm{AND} \\
\mathrm{z} - \mathrm{NB527} > 0.5
\end{split}
\end{eqnarray}
for $z=3.3$ LAEs, 
\begin{eqnarray}
\begin{split}
\mathrm{g} > \mathrm{g_{2\sigma}}~\mathrm{AND}~\mathrm{NB718} < \mathrm{NB718_{5\sigma}}~\mathrm{AND} \\
\mathrm{ri} - \mathrm{NB718} > \mathrm{max}(0.7, (\mathrm{ri}-\mathrm{NB718})_\mathrm{3\sigma}) ~\mathrm{AND} \\
\mathrm{r} - \mathrm{i} > 0.8
\label{eq:z49}
\end{split}
\end{eqnarray}
for $z=4.9$ LAEs, where ri is calculated by the linear combination of the fluxes in the r band, $f_\mathrm{r}$, and those of the i band, $f_\mathrm{i}$, as $f_\mathrm{ri} = 0.3f_\mathrm{r} + 0.7f_\mathrm{i}$, 
\begin{eqnarray}
\begin{split}
\mathrm{g} > \mathrm{g_{2\sigma}}~\mathrm{AND} ~\mathrm{NB816} < \mathrm{NB816_{5\sigma}}~\mathrm{AND} \\
\mathrm{i} - \mathrm{NB816} > \mathrm{max}(1.2, (\mathrm{i}-\mathrm{NB816})_\mathrm{3\sigma})~\mathrm{AND} \\
((\mathrm{r} > \mathrm{r_{3\sigma}}) ~\mathrm{OR}~
(\mathrm{r} < \mathrm{r_{3\sigma}} ~\mathrm{AND}~
\mathrm{r} - \mathrm{i} > 1.0))
\end{split}
\end{eqnarray}
for $z=5.7$ LAEs, 
\begin{eqnarray}
\begin{split}
\mathrm{g} > \mathrm{g_{2\sigma}}~\mathrm{AND}~
\mathrm{r} > \mathrm{r_{2\sigma}}~\mathrm{AND} \\
\mathrm{NB921} < \mathrm{NB921_{5\sigma}}~\mathrm{AND} \\
\mathrm{z} - \mathrm{NB921} > \mathrm{max}(1.8, (\mathrm{z}-\mathrm{NB921})_\mathrm{3\sigma})~\mathrm{AND} \\
((\mathrm{z} > \mathrm{z_{3\sigma}}) ~\mathrm{OR}~
(\mathrm{z} < \mathrm{z_{3\sigma}} ~\mathrm{AND}~
\mathrm{i} - \mathrm{z} > 1.0))
\end{split}
\end{eqnarray}
for $z=6.6$ LAEs, 
\begin{eqnarray}
\begin{split}
\mathrm{g} > \mathrm{g_{2\sigma}}~\mathrm{AND}~ 
\mathrm{r} > \mathrm{r_{2\sigma}}~\mathrm{AND}~
\mathrm{i} > \mathrm{i_{2\sigma}}~\mathrm{AND} \\
\mathrm{NB973} < \mathrm{NB973_{5\sigma}}~\mathrm{AND} \\
((\mathrm{z} > \mathrm{z_{3\sigma}}) ~\mathrm{OR}~
(\mathrm{z} < \mathrm{z_{3\sigma}} ~\mathrm{AND}~
\mathrm{z} - \mathrm{y} > 2.0))~\mathrm{AND} \\
\mathrm{y} - \mathrm{NB973} > 0.7
\end{split}
\end{eqnarray}
for $z=7.0$ LAEs, 
and
\begin{eqnarray}
\begin{split}
\mathrm{g} > \mathrm{g_{2\sigma}}~\mathrm{AND}~
\mathrm{r} > \mathrm{r_{2\sigma}}~\mathrm{AND}~
\mathrm{i} > \mathrm{i_{2\sigma}}~\mathrm{AND}~
\mathrm{z} > \mathrm{z_{2\sigma}}~\mathrm{AND} \\
\mathrm{NB1010} < \mathrm{NB1010_{5\sigma}}~\mathrm{AND}~ \mathrm{y} - \mathrm{NB1010} > 1.9 \hspace{1.0cm}
\end{split}
\end{eqnarray}
for $z=7.3$ LAEs.

\begin{deluxetable*}{lcccccccccccccc}
\tablecolumns{15}
\tabletypesize{\scriptsize}
\tablecaption{Survey Area
\label{table:survey}}
\tablehead{
\colhead{Field} & \colhead{Area NB387} & \colhead{Area NB527} & \colhead{Area NB718} & \colhead{Area NB816} & \colhead{Area NB921} & \colhead{Area NB973} & \colhead{Area NB1010} \\
 & (deg$^2$) & (deg$^2$) & (deg$^2$) & (deg$^2$) & (deg$^2$) & (deg$^2$) & (deg$^2$)}
\startdata
 UD\Add{-}COSMOS & 2.23 & 1.82 & 1.79 & 2.04 & 2.03 & 1.75 & 1.79 \\
 UD\Add{-}SXDS & \nodata & \nodata & \nodata & 1.72 & 1.70 & \nodata & 1.79 \\
 D\Add{-}COSMOS & 5.03 & \nodata & \nodata & 4.73 & 4.77 & \nodata & \nodata \\
 D\Add{-}SXDS & 4.63 & \nodata & \nodata & 4.33 & 4.36 & \nodata & \nodata \\
 \Add{D-}DEEP23 & 6.20 & \nodata & \nodata & 5.77 & 5.80 & \nodata & \nodata \\
 \Add{D-}ELAISN1 & 6.35 & \nodata & \nodata & 5.71 & 5.72 & \nodata & \nodata \\
 \hline
Total & 24.43 & 1.82 & 1.79 & 24.03 & 24.38 & 1.75 & 3.58 \\\hline
\enddata
\label{tab:surveyarea}
\tablecomments{NB816 and NB921 cover the entire D and UD layers of the HSC-SSP survey. NB387 covers those fields except for the UD-SXDS field when combined with the CHORUS survey data. NB1010 covers the UD\Add{-}COSMOS and UD\Add{-}SXDS fields. NB527, NB718, and NB973 filters from the CHORUS survey cover the UD\Add{-}COSMOS field.}
\end{deluxetable*}

These color criteria correspond to the rest-frame Ly$\alpha$ equivalent width (EW) of  $\sim10-20$ {\AA} for the case of a flat UV continuum and IGM attenuation \citep{1999ApJ...514..648M}. Note that we applied a stricter selection criterion for $z=6.6$ sample than the one used in \cite{Kikuta23} to match the EW limit with that of previous studies \citep{2006PASJ...58..313S,Ouchi10,Konno18}. The size of LAE samples for $z=2.2$, 3.3, 4.9, 5.7, 6.6, 7.0, and 7.3 are 6995, 4641, 726, 6124, 451, 18, and 5, respectively. In total, our LAE catalog consist of 18960 objects. 

\subsection{Spectroscopically Confirmed Sample}
In addition to the photometrically selected sample, \Add{\cite{Kikuta23} also constructed spectroscopically confirmed LAE samples based on the subsample consisting of spectroscopically confirmed LAE candidates (hereafter, spec subsample).} \cite{Kikuta23} present all the spectroscopically confirmed LAE candidates determined from the cross-match between our LAE sample and all available spectroscopic data and literature \citep[][ and Ouchi et al. in preparation]{Ouchi08,Lilly09,Ouchi10,Coil11,Mallery12,Masters12,LeFevre13,Shibuya14,Kriek15,Liske15,Sobral15,Momcheva16,Tasca17,
Hu17,Jiang17,Harikane18,Hasinger18,Jiang18,Scodeggio18,Shibuya18a,Shibuya18b,Harikane19,Lyke20,Ning20,Ono21,Ning22}. The spectroscopically confirmed sources in our photometric LAE sample \Add{are} summarized in Table 4 of \cite{Kikuta23}. \Add{More comprehensive spectroscopic follow-ups of our LAE samples will be completed using multi-object spectroscopy with the upcoming Subaru/Prime Focus Spectrograph (PFS) observations \citep{Greene22}.}\par

\section{Luminosity Function} \label{sec:lf}

\subsection{Detection Completeness}
\begin{figure}[htbp]
      \begin{minipage}[t]{\hsize}
        \centering
        \includegraphics[width=\linewidth]{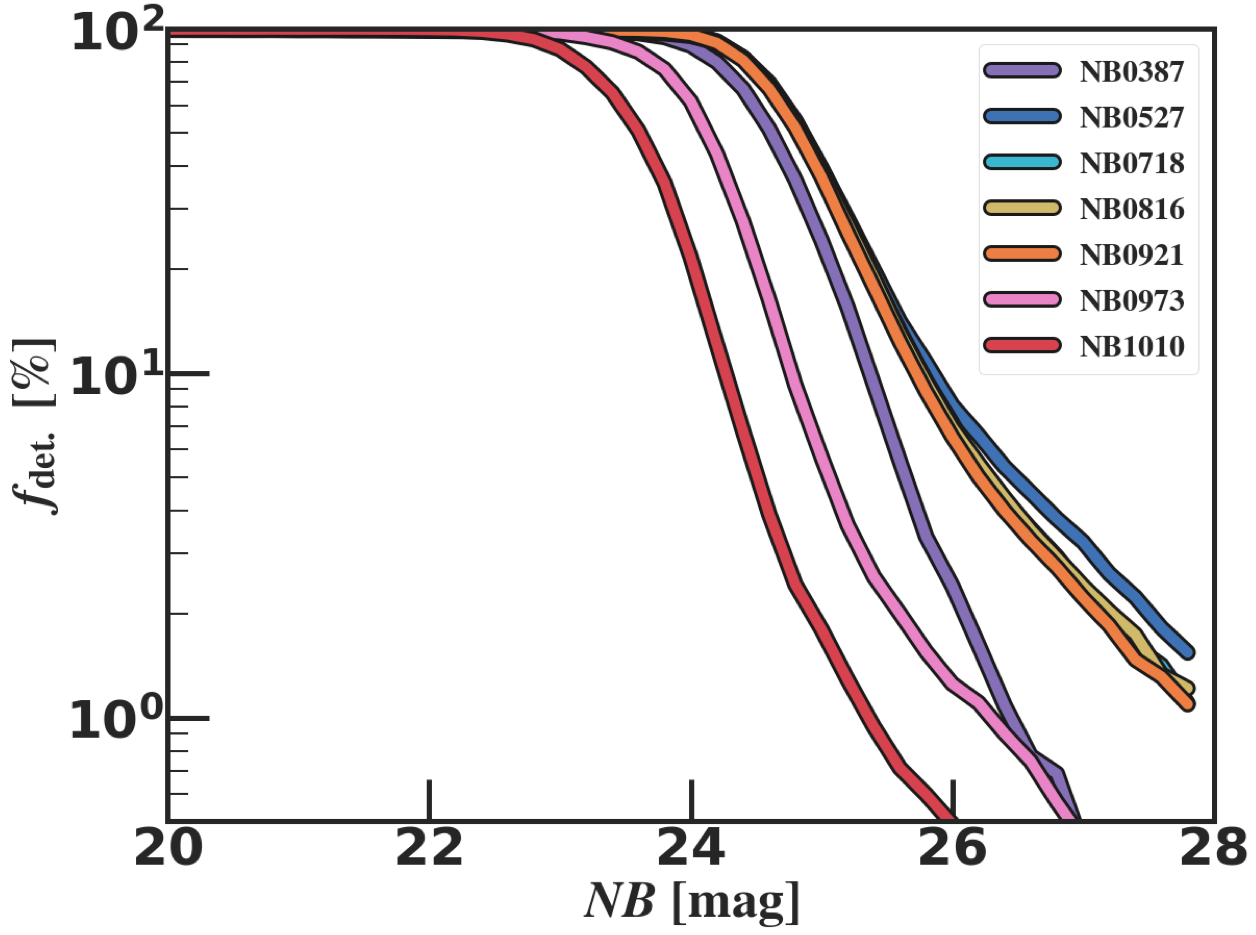}
      \end{minipage} 
      \begin{minipage}[t]{\hsize}
        \centering
        \includegraphics[width=\linewidth]{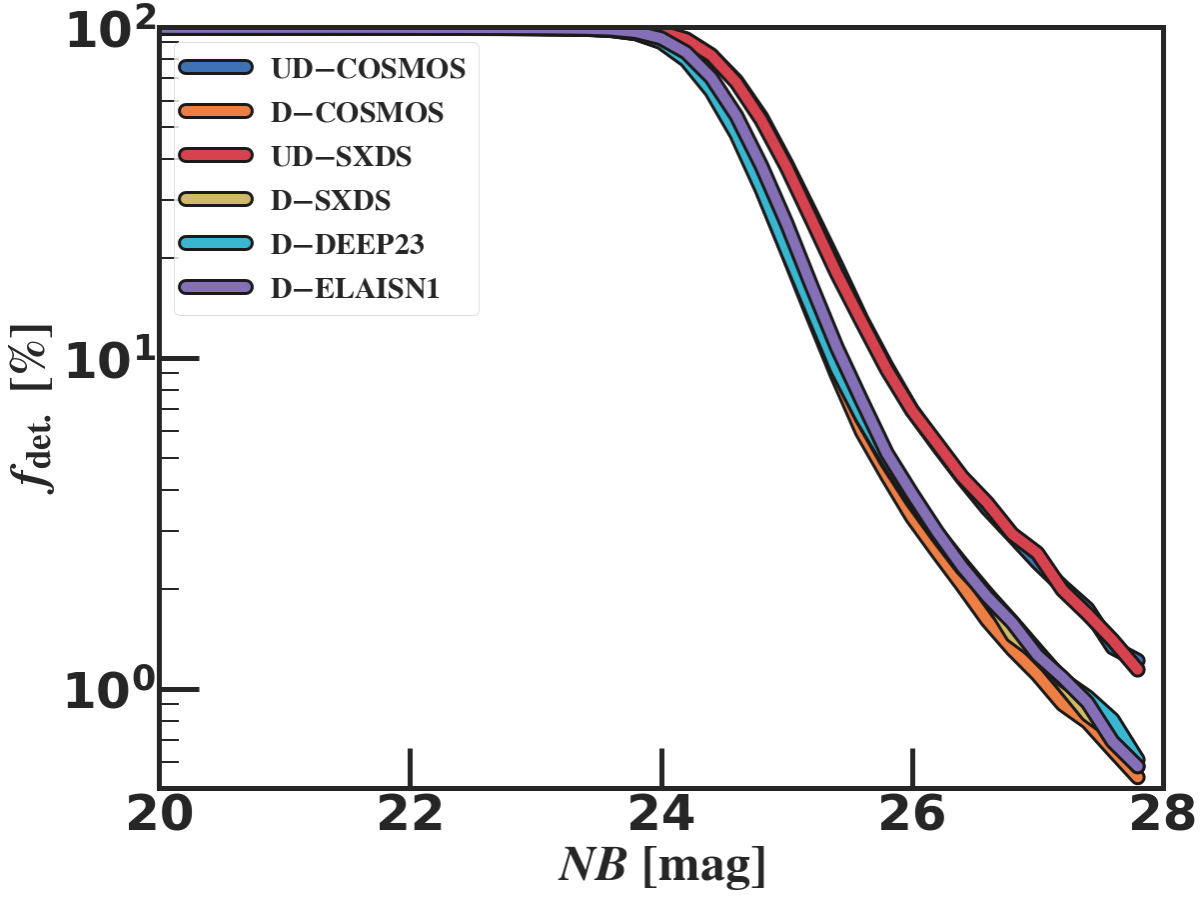}
      \end{minipage}
      \caption{(Top) Detection completeness, $f_{\rm det}$ of our narrow band images taken with Subaru/HSC in UD\Add{-}COSMOS field. The x-axis (i.e., $NB~[\rm mag]$) represent the total narrow-band magnitude. The purple, blue, pale-blue, yellow, orange, pink, and red lines represent $f_{\rm det}$ at each narrow-band magnitude for $NB387$, $NB527$, $NB718$, $NB816$, $NB921$, $NB973$, and $NB1010$ images, respectively. (Bottom) $f_{\rm det}$ of our $NB816$. The blue, orange, red, yellow, pale-blue, and purple lines represent $f_{\rm det}$ in UD\Add{-}COSMOS, D\Add{-}COSMOS, UD\Add{-}SXDS, D\Add{-}SXDS, \Add{D-}DEEP23, and \Add{D-}ELASIN1 fields, respectively.}
\label{nb0387_comp} 
\end{figure}

\Add{To determine the number density of LAEs,} we estimate the detection completeness of our sample in the similar way as \cite{Konno18}. We estimate the detection completeness of narrowband images as a function of narrowband total magnitude. As \cite{Kikuta23} mention, the detection completeness \Add{is spatially} different even in the same survey field. Because the original HSC images are divided into ``patches" with 4100 or 4200 pixels on a side ($\sim11.5'$), we determine the detection completeness for each patch to account for the spatial variance in image depth. We insert 1000 mock LAE objects into each patch image. We then detect LAEs in the same manner as LAEs are detected in \cite{Kikuta23}. We randomly generate mock LAE\Add{s} with \Add{a} surface brightness profile following \Add{a} radial profile probability distribution described in \cite{Shibuya19}. We then derive the area-averaged detection completeness for each survey field. We calculate the detection completeness at the narrow-band magnitude range of 20.0 to 27.8 with a step size of 0.2. In Figure \Add{\ref{nb0387_comp}}, we show the detection completeness as a function of narrow-band total magnitude for different narrow-bands in UD-COSMOS field and for $NB387$ in different survey fields, respectively.

\subsection{Contamination} \label{lf:cont}
We consider contamination by low-$z$ interlopers and other objects in our \Add{photometric} LAE samples. Because we have not conducted spectroscopic follow-ups on our new LAE photometric candidates, \Add{we refer to the previous work and the cross-match between our sample and spectroscopically confirmed objects conducted in \cite{Kikuta23}. For low-z LAE samples from $NB0387$ and $NB0527$ imaging data, \cite{Kikuta23} suggested that the foreground (i.e., $z~0$) line emitters would have small contributions to our LAE samples due to the small volume size probed with narrow-band at low redshift (i.e., $z~0$). For example, \cite{Kikuta23} discuss that the contamination by $z=0.03$ {\OII} emitter to $z=2.2$ LAE sample would be negligible as the survey volume for $z=0.03$ would only be 0.2\% of that for $z=2.2$ LAEs.} \Add{For the high-z samples}, we refer to the previous results shown in \cite{Shibuya18b} for the contamination fraction, $f_{\rm cont}$. \cite{Shibuya18b} conducted spectroscopic follow-up observations of the LAE candidate that are selected in an almost identical manner as it has been done in \cite{Kikuta23}. After the selection, \cite{Shibuya18b} conducted spectroscopic follow-up observations for 81 LAE candidates at $z=5.7$ and 6.6. Out of the 28 (53) LAE candidates at $z=5.7$ (6.6), 4 (4) are identified as low-$z$ interlopers. Based on these results, \cite{Konno18} estimate $f_{\rm cont}$ as $f_{\rm cont}=4/28\simeq14\%$ ($4/53\simeq8\%$) for $z=5.7$ (6.6) LAE candidates. \cite{Shibuya18b} also estimate $f_{\rm cont}$ for the bright ($NB<24$ mag) LAE candidates and found $f_{\rm cont}$ of $z=5.7$ (6.6) is $f_{\rm cont}=4/12\simeq33\%$ ($=1/6\simeq17\%$). \Add{We do note, however, that \cite{Kikuta23} have found that all bright contaminants at $z=5.7$ and 6.6 reported in \cite{Shibuya18b} are successfully omitted during the blind selection processes.} Based on these \Add{findings}, we \Add{do not apply contamination correction in our work as we expect small contamination fraction in our sample and the contaminant would have negligible impact on our statistical results otherwise mentioned.}
\Add{The confirmation of contamination fraction will be addressed in the upcoming Subaru/PFS spectroscopic observations.}\\

\begin{figure*}[htbp]
    \centering
    \includegraphics[width=0.9\linewidth]{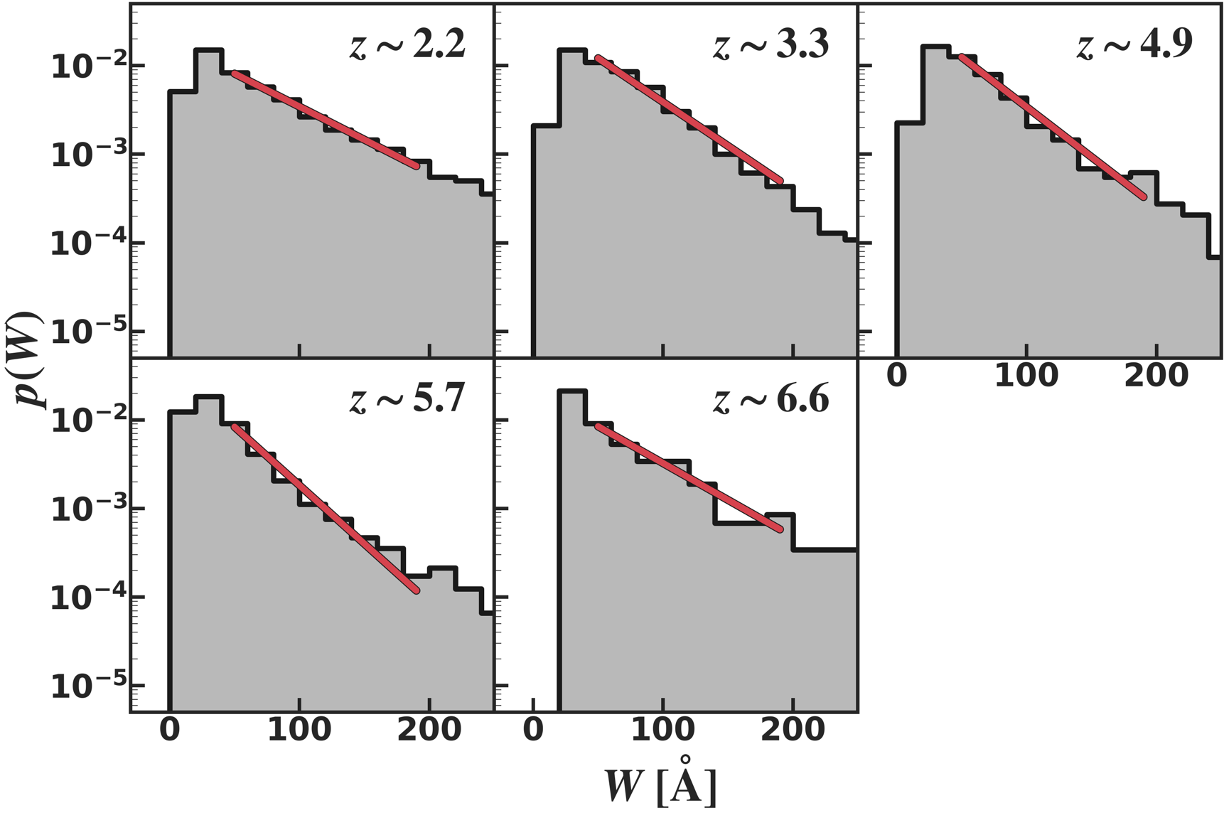}
    \caption{Photometrically derived rest-frame equivalent width distribution of our LAE sample at $z\sim2.2$, 3.3, 4.9, 5.7, and 6.6. Grey histogram represent the normalized probability density $p(EW_0)$ of our LAE sample. The red solid lines in each panel represent the best-fit exponential distribution fitted in the range of $EW_0=40-200$ {\AA}.}
    \label{EW_dist}
\end{figure*}

\begin{figure*}[htbp]
    \centering
    \includegraphics[width=0.95\linewidth]{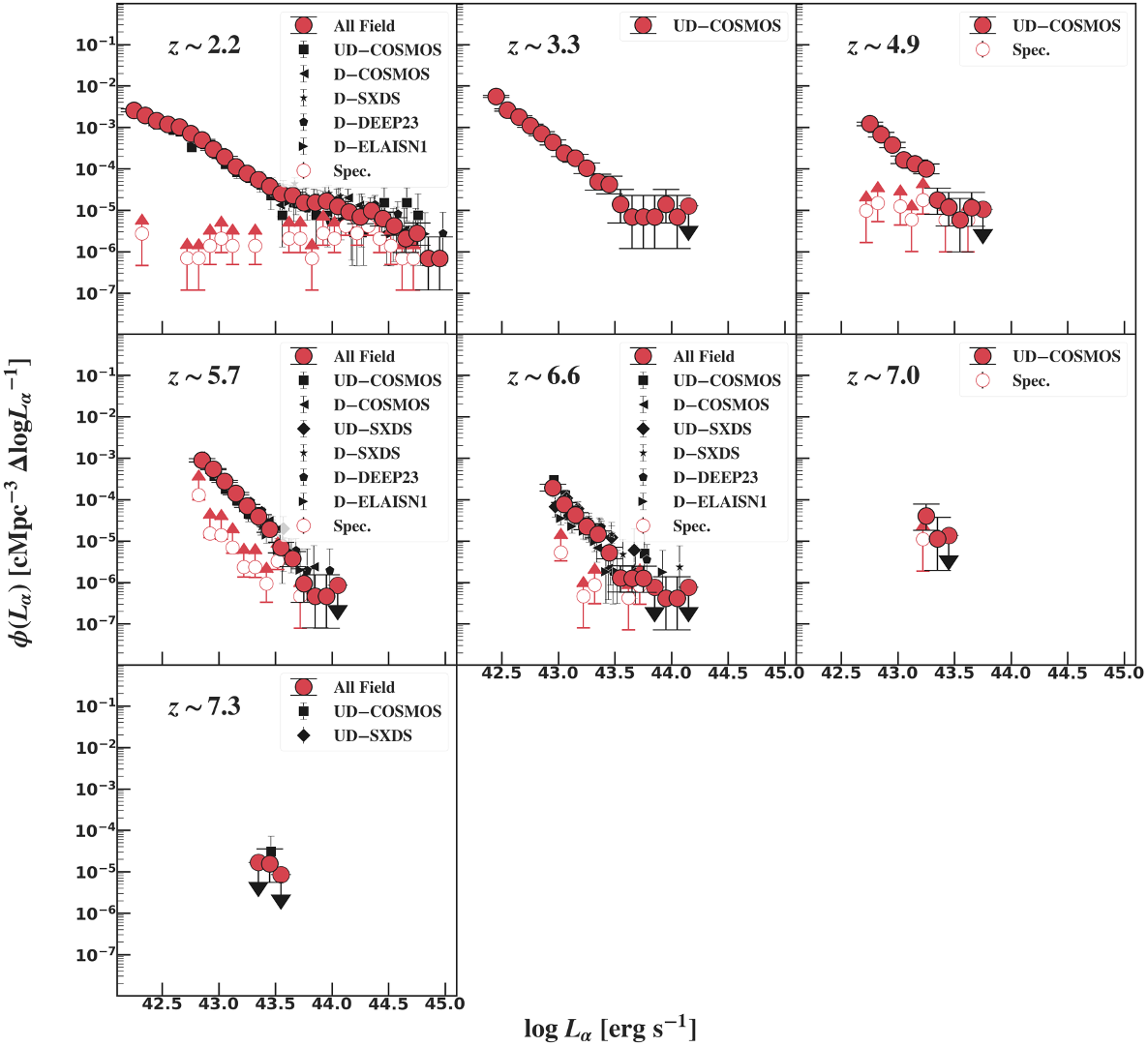}
    \caption{The luminosity function of this work at $z=2.2$, 3.3, 4.9, 5.7, 6.6, 7.0, and 7.3. The \Add{red} filled circle and bars represent the number densities and their errors at each redshift using the data from all survey field. \Add{The black} squares, left tipped triangles, diamonds, stars, pentagons, and right tipped triangles represent the number density calculated for the UD\Add{-}COSMOS, D\Add{-}COSMOS, UD\Add{-}SXDS, D\Add{-}SXDS, \Add{D-}DEEP23, and \Add{D-}ELAISN1 fields, respectively. The red \Add{open} circle with upper-side arrow represents the number density and the lower limit based on the spectroscopic identified sources in our photometric sample. The spectroscopic confirmation of objects in our photometric LAE samples are based on \cite{Ouchi08}, \cite{Lilly09}, \cite{Ouchi10}, \cite{Coil11}, \cite{Mallery12}, \cite{Masters12}, \cite{LeFevre13}, \cite{Shibuya14}, \cite{Kriek15}, \cite{Liske15}, \cite{Sobral15}, \cite{Momcheva16}, \cite{Tasca17},
    \cite{Hu17}, \cite{Jiang17}, \cite{Harikane18}, \cite{Hasinger18}, \cite{Jiang18}, \cite{Scodeggio18}, \cite{Shibuya18a}, \cite{Shibuya18b}, \cite{Harikane19}, \cite{Lyke20}, \cite{Ning20}, \cite{Ono21}, \cite{Ning22}, and Ouchi et al. in preparation \citep[see details in ][]{Kikuta23}. All the number density presented in this figure are corrected for the detection completeness.}
    \label{lf_data}
\end{figure*}

\subsection{Ly$\alpha$ Luminosity Functions}
We calculate the Ly$\alpha$ luminosity function (LF) in the so-called ``classical" method described in \cite{Konno18}. In the classical method, we assume Ly$\alpha$ emission with a delta-function line profile located at the center of the NB filter. We also assume that the UV continuum flux ($f_c$) is constant (i.e., $f_c\propto\nu^0$) and the complete attenuation by IGM at the blueward of the Ly$\alpha$ emission. With this assumption, the rest-frame equivalent width ($EW_{0}$) and flux of Ly$\alpha$ ($f_\alpha$) can be expressed using the narrow- and broad-band magnitude ($m_{\rm NB}$ and $m_{\rm BB}$, respectively) as follows:

\begin{eqnarray}
    EW_0 =& \frac{\lambda_\alpha^2}{c(1 + z_{\rm Ly\alpha})}\frac{f_{\rm NB}/C_{\rm NB} - {f_{\rm BB}}/C_{\rm BB}}{A_{\rm NB}/C_{\rm NB} - A_{\rm BB}/C_{\rm BB}},\\
    f_\alpha =& \frac{f_{\rm NB}/C_{\rm NB} - f_{\rm BB}/C_{\rm BB}}{A_{\rm NB}/C_{\rm NB} - A_{\rm BB}/C_{\rm BB}}.
\end{eqnarray}
Here, $f_{\star}$, $C_{\star}$ and $A_{\star}$ ($\star=$NB or BB) are

\begin{eqnarray}
    f_{\rm \star} &= 10^{-0.4(m_\star + 48.60)}, \\
    A_{\rm \star} &= \frac{T(\nu_\alpha)/\nu_{\alpha}}{\int^{\infty}_{0}d\nu T_{\rm \star}(\nu)/\nu}, \\
    C_{\rm \star} &= \frac{\int^{\nu_\alpha}_{0}d\nu T_{\star}(\nu)/\nu}{\int^{\infty}_{0}d\nu T_{\rm \star}(\nu)/\nu},
\end{eqnarray}
where $T_\star(\nu)$ is a filter transmission at the frequency $\nu$ for narrow-band or broad-band filter. \Add{We present the rest-frame equivalent width distribution at redshift between $z=2.2$ to 6.6 in Figure \ref{EW_dist}.}\par
We calculate the luminosity function by binning the LAE sample by Ly$\alpha$ luminosity with the binning size of $\Delta \log L_\alpha=0.1$ and correction for the detection completeness. We calculate the survey volumes covered by the survey field and the redshift ranges corresponding to the full width at half maximum (FWHM) of the NB filters. We calculate the error of the luminosity function at each luminosity bin assuming the Poisson statistics. Because the 1$\sigma$ derived analytically from the Poisson distribution (i.e., the square root of the number counts) underestimates the actual 1$\sigma$ uncertainties for the small number counts, we instead use the values presented in Table\Add{s} 1 and 2 of \cite{Gehrels86} for upper and lower 1$\sigma$ uncertainties, respectively. We define the limiting Ly$\alpha$ luminosity based on the NB magnitude corresponding to the 50\% of detection completeness. We calculate the luminosity function above the limiting Ly$\alpha$ luminosity in each survey field. We also calculate the average luminosity function by combining the results in all fields at each redshift. We present the luminosity function values for all fields at each redshift in Table \ref{table:num_den}. In Figure \ref{lf_data}, we present both the field-to-field and all-field luminosity functions for each redshift LAE sample above the limiting Ly$\alpha$ luminosity. \par
We also present the luminosity function based on the spectroscopically confirmed objects (hereafter, spec Ly$\alpha$ luminosity functions). \Add{The spec Ly$\alpha$ luminosity functions are calculated in the same way as the photometric sample, but we only use the spectroscopically confirmed photometric LAE candidates.} We show spec Ly$\alpha$ luminosity functions as lower limits in Figure \ref{lf_data} as our spec LAE samples only represent portions of the total photometric LAE candidates.\par  

\subsection{Fitting Luminosity Functions}
We fit the luminosity function ($\phi$) across different luminosity bins with several analytic functions. We first fit with the Schechter function ($\phi_{\rm Sch}$), which is given by
\begin{equation}
    \phi_{\rm Sch}(L)d\Add{\log}L = (\ln 10)\phi^\star{\left(\frac{L}{L^\star}\right)}^{\alpha + 1}\exp\left(-\frac{L}{L^\star}\right)d\log L.
    \label{Schechter}
\end{equation}
Here, $\phi^\star$, $L^\star$, and $\alpha$ represents the characteristic number density, characteristic luminosity, and the faint-end slope of the luminosity function, respectively. At the higher redshifts, UV luminosity function starts to show a number density excess at the bright-end that cannot be well explained by the Schechter function \citep[e.g.,][]{Harikane22}. \cite{Harikane22} show that a luminosity function with such a bright-end excess can be well expressed with the double power law (DPL) function ($\phi_{\rm DPL}$). Double power law function is given by
\begin{equation}
\begin{aligned}
    \phi_{\rm DPL}(L)d\Add{\log}L &= (\ln 10)\phi^\star\\
    & \times{\left[\left(\frac{L}{L^\star}\right)^{-\alpha - 1} + \left(\frac{L}{L^\star}\right)^{-\beta - 1}\right]}^{-1}d\log L,
\end{aligned}
\end{equation}
where $\beta$ is an additional parameter called the bright-end slope and $\alpha > \beta$. \par 
We infer the posterior probability distribution functions of the luminosity function parameters. We adopt \Add{a} uniform, bounded prior probability distribution for $\log \phi^\star$, $\log L^\star$, $\alpha$, and $\beta$. We explore the parameter space of $\log \phi^\star/\mathrm{cMpc}^{-3}=-8.0$ to 0.0, $\log L^\star/\mathrm{erg}~\mathrm{s}^{-1}=42.0$ to 44.0, $\alpha$=$-4.0$ to $-1.0$, and $\beta$=$-8.0$ to max($\alpha$, $-1.0$).
For the likelihood function ($\mathcal{L}$), we adopt a gaussian functional form as follows:
\begin{equation}
\begin{aligned}
    \ln \mathcal{L} (\vec{\theta}) = &-\frac{1}{2}\sum_i\left(\frac{\phi_{{\rm obs},i} - \phi_{\rm mod}(L_i;\vec{\theta})}{\sigma_{{\rm obs},i}}\right)^{2}\\
    &-\frac{1}{2}\ln(2\pi\sigma_{{\rm obs},i}^2).
    \label{lumlike}
\end{aligned}
\end{equation}
In Equation \ref{lumlike}, $\vec{\theta}$, $L_i$, $\phi_{\rm mod}$, and $\phi_{{\rm obs},i}$ ($\sigma_{{\rm obs},i}$) represent luminosity function parameter sets, luminosity in the $i$-th luminosity bin, fitting function, and the number density (uncertainty) of the observed luminosity function in $i$-th luminosity bin, respectively. To obtain posterior probability distribution functions, we conduct Markov Chain Monte Carlo (MCMC) method. We use an affine invariant MCMC ensemble sampler {\sc emcee} \citep{emcee}. \par
Because Ly$\alpha$ LFs from this work mainly cover the luminosity range of $\log L_\alpha/{\rm erg~s^{-1}}=42.4-44.0$, we use the literature value to complement faint/bright end of LFs. For $z=2.2$, we use value from \cite{Cassata11} and \cite{Zhang21} for the luminosity range of $\log L_\alpha/{\rm erg~s^{-1}}=41.3-42.0$ and \Add{$43.3-45.4$}\Add{, respectively.} For the bright-end of the $z=2.2$ luminosity function, we use the value from \cite{Zhang21} instead of those from this work because \cite{Zhang21} derive luminosity function \Add{from purely spectroscopic sample with a sample selection based on spectroscopy to isolate LAE sample from the foreground bright {\OII} emitters.} 
At $z=3.3$, we use the values from \cite{Drake17} in the luminosity range of $\log L_\alpha/{\rm erg~s^{-1}}=41.2-42.0$. At $z=4.9$, we use the values from \cite{Drake17} in the luminosity range of $\log L_\alpha/{\rm erg~s^{-1}}=41.2-42.4$. At $z=5.7$, we use the values from \cite{Drake17} and \cite{Ouchi08} in the luminosity range of $\log L_\alpha/{\rm erg~s^{-1}}=41.2-42.0$ and $42.5-42.7$, respectively. At $z=6.6$, we use \cite{Ouchi10} values in the luminosity range of $\log L_\alpha/{\rm erg~s^{-1}}=42.5-42.7$. At $z=7.0$, we use the values from \cite{Ota17} in the luminosity range of $\log L_\alpha/{\rm erg~s^{-1}}=42.7-43.1$, respectively. We use results from \cite{Konno14} at the luminosity range of $\log L_\alpha/{\rm erg~s^{-1}}42.5-42.9$ for $z=7.3$.\par
Some literature suggests the existence of significant AGN component on the brighter-end of Ly$\alpha$ luminosity function at $z\sim2-3$ \cite[e.g., ][]{Konno16, Sobral18}. In Figure \ref{lf_data}, we do see the broken power-law features around $\log L_\alpha\simeq43.5$ at $z=2.2$, indicating non-negligible contribution from AGN to Ly$\alpha$ luminosity function. We do also see tentative broken power-law features around $\log L_\alpha\simeq43.5$ for $z=3.3$ and 4.9 Ly$\alpha$ LF in Figure \ref{lf_data}. For this reason, we consider \Add{an} AGN component in the Ly$\alpha$ LFs at $z\leq4.9$. We model \Add{the} AGN component by the DPL function. We refer to \Add{a} fitting model that assumes the galaxy component is a DPL (or Schechter) function as DPL + DPL (or Schechter + DPL). To distinguish the parameters between components, we add subscripts `gal' and `agn' to luminosity function parameters for galaxy and AGN components, respectively. \Add{We adopt the same uniform prior for the luminosity function parameters of galaxy components as the single component cases. For the AGN components, we adopt an uniform, bounded prior probability distribution in the following parameter spaces: $-10.0\leq\log \phi^\star_{\rm agn}/\mathrm{cMpc}^{-3}\leq{\rm max}(-3.0,~\log \phi^\star_{\rm gal}/\mathrm{cMpc}^{-3}$), ${\rm max}(43.5,~\log L^\star_{\rm gal}/\mathrm{erg}^{-1})\leq\log L^\star/\mathrm{erg}^{-1}\leq45.0$, $-4.0\leq\alpha_{\rm agn}\leq1.0$, and $-8.0\leq\beta_{\rm agn}\leq {\rm max}(\alpha_{\rm agn},~-1.0).$}

\begin{figure*}[htbp]
    \includegraphics[width=0.9\linewidth]{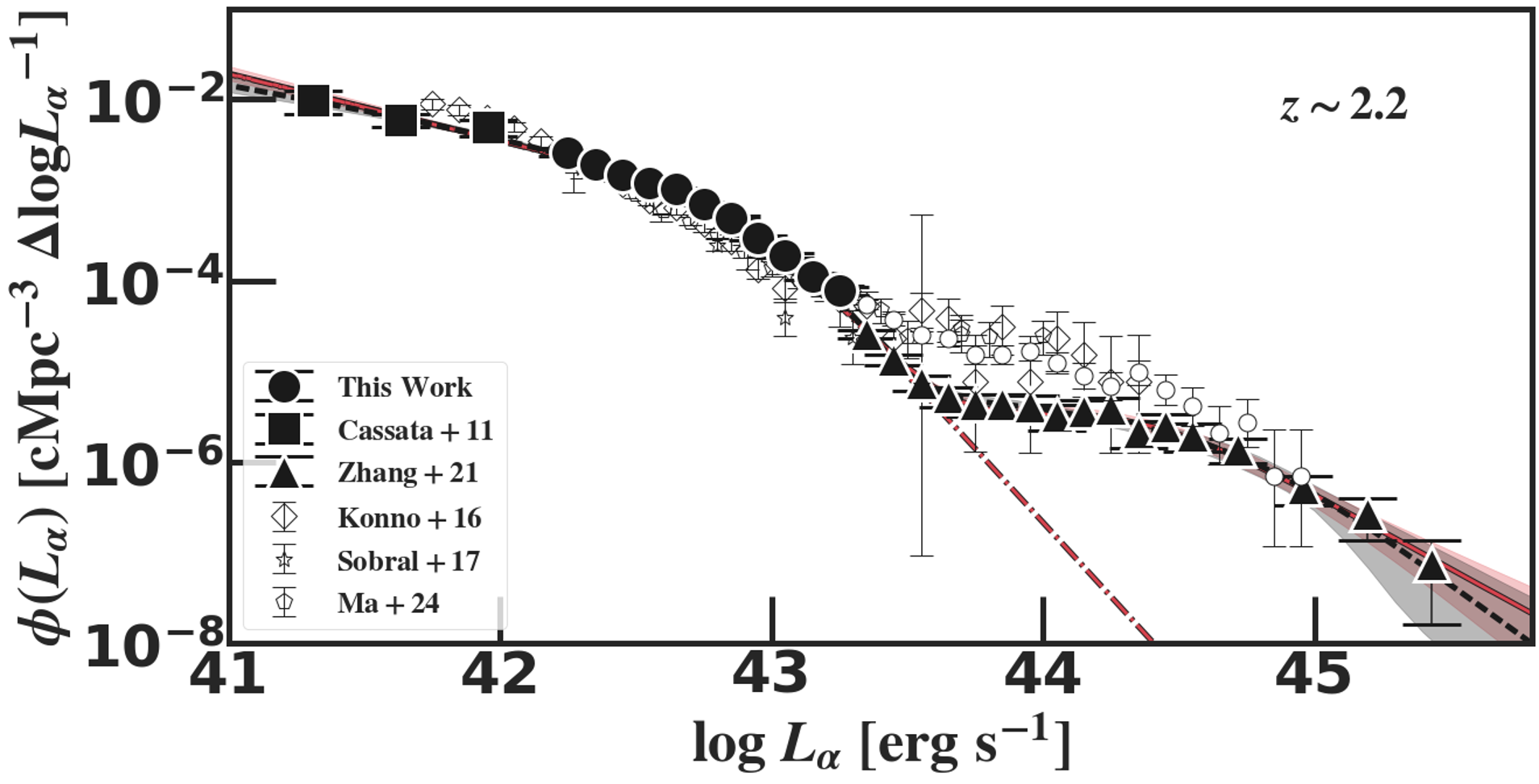}
    \caption{The fitting result for Ly$\alpha$ LF at $z=2.2$. The circles, squares, upward triangles, diamonds, \Add{stars, and pentagons} with bars represent the number densities and corresponding errorbars from this work, \cite{Cassata11}, \cite{Zhang21}, \cite{Konno16}, \cite{Sobral17}, and \cite{Ma24}, respectively. Filled (open) marks are \Add{(not)} used to fit the Ly$\alpha$ LF functions. The median of the fitted DPL + DPL and Schechter + DPL function at each luminosity is shown in the red solid and black dashed lines, respectively. \Add{The} 68-th percentile range for the fitted DPL + DPL and Schechter + DPL functions at each luminosity is shown by red and gray shades, respectively. \Add{The red dotted-dashed line represent the median of the galaxy component for DPL + DPL function.}}
    \label{fit_lf_z2p2}
\end{figure*}

\begin{figure*}[htbp]
    \includegraphics[width=0.9\linewidth]{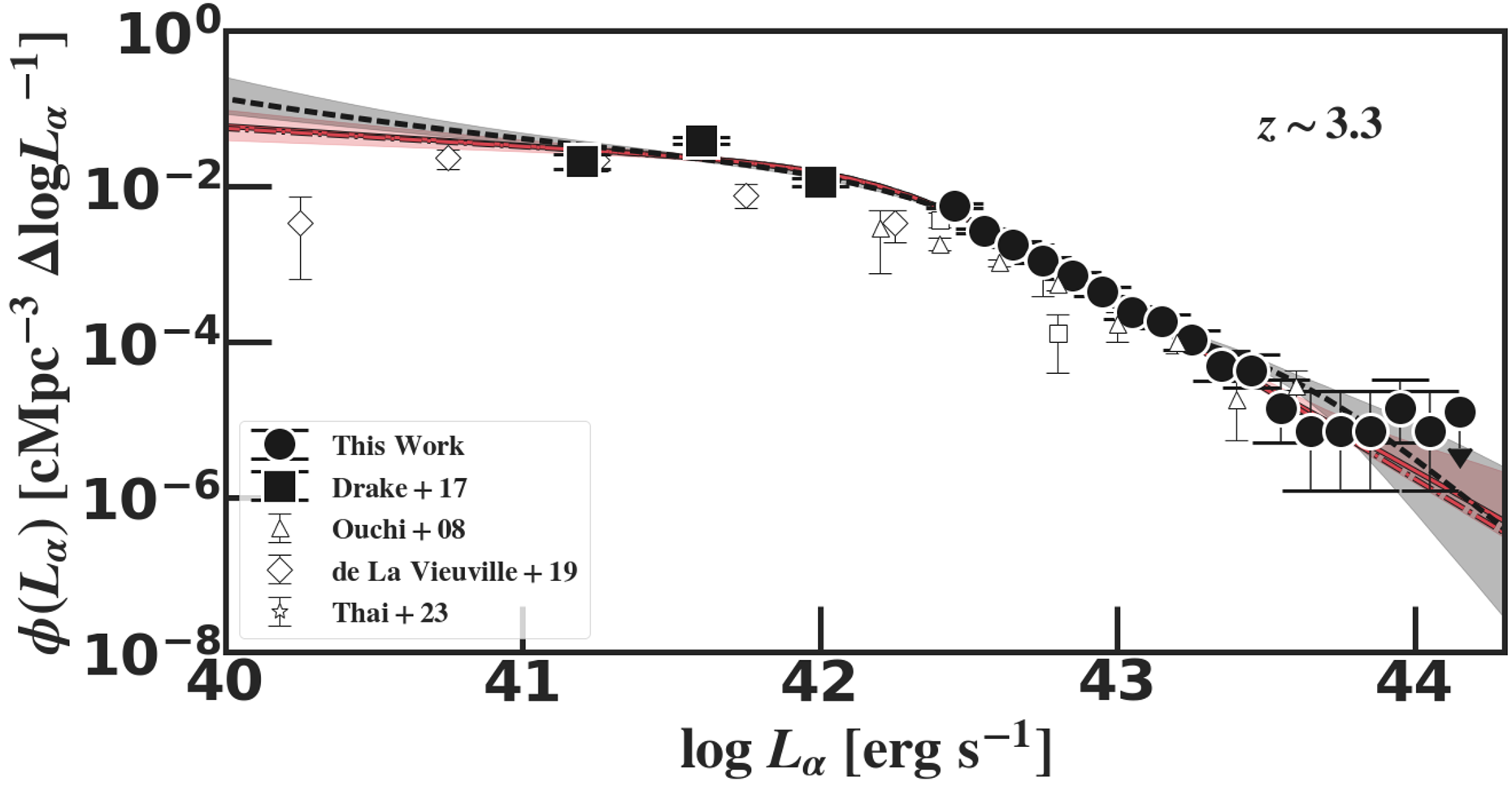}
    \caption{The fitting result for Ly$\alpha$ LF at $z=3.3$. The circles, squares, upward triangles, diamonds, and stars with bars represent the the number densities and corresponding errorbars from this work, \cite{Ouchi08}, \cite{Drake17}, \cite{LadeVieuville19}, and \cite{Thai23}, respectively. Filled (open) marks are \Add{(not)} used to fit the Ly$\alpha$ LF functions. The other symbols represent the same as in the Figure \ref{fit_lf_z2p2}.}
    \label{fit_lf_z3p3}
\end{figure*}

\begin{figure*}[htbp]
    \includegraphics[width=\linewidth]{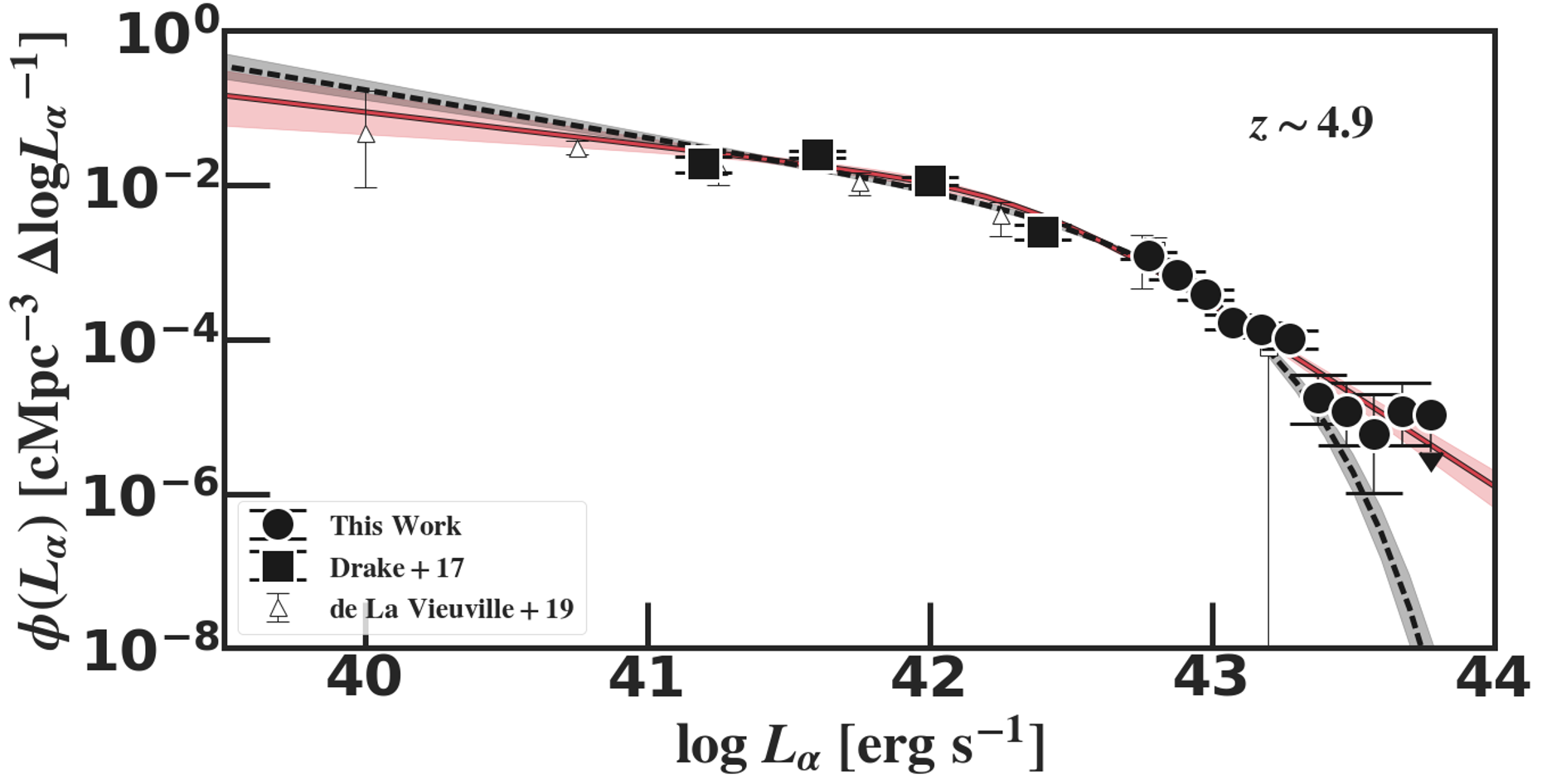}
    \caption{The fitting result for Ly$\alpha$ LF at $z=4.9$. The circle, squares, and upward triangles with bars represent the number densities and corresponding errorbars from this work, \cite{Drake17}, and \cite{LadeVieuville19}, respectively. Filled (open) marks are \Add{(not)} used to fit the Ly$\alpha$ LF functions. The median of the fitted DPL and Schechter function at each luminosity is shown in the red solid and black dashed lines, respectively. \Add{The} 68-th percentile range for the fitted DPL and Schechter functions at each luminosity is shown by red and gray shades, respectively.}
    \label{fit_lf_z4p9}
\end{figure*}

\begin{figure*}[htbp]
    \includegraphics[width=\linewidth]{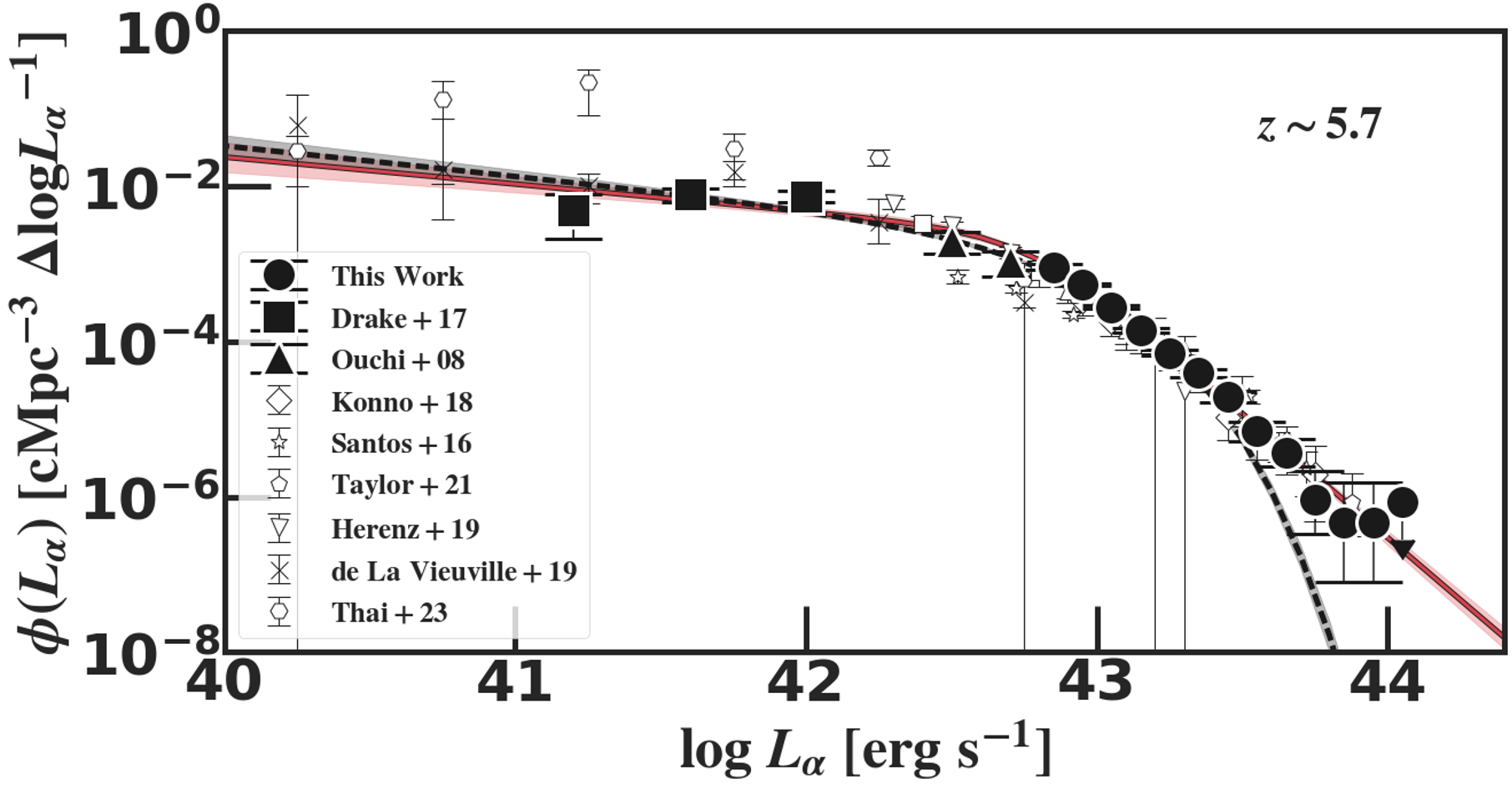}
    \caption{The fitting result for Ly$\alpha$ LF at $z=5.7$. The circle, squares, upward triangles, diamonds, stars, pentagons, downward triangles, cross, and hexagons with bars represent the number densities and corresponding errorbars from this work, \cite{Drake17}, \cite{Ouchi08}, \cite{Konno18}, \cite{Santos16}, \cite{Taylor21}, \cite{Herenz19}, \cite{LadeVieuville19}, and \cite{Thai23}, respectively. Filled (open) marks are \Add{(not)} used to fit the Ly$\alpha$ LF functions. The other symbols represent the same as in the Figure \ref{fit_lf_z4p9}.}
    \label{fit_lf_z5p7}
\end{figure*}

\begin{figure*}[htbp]
    \includegraphics[width=\linewidth]{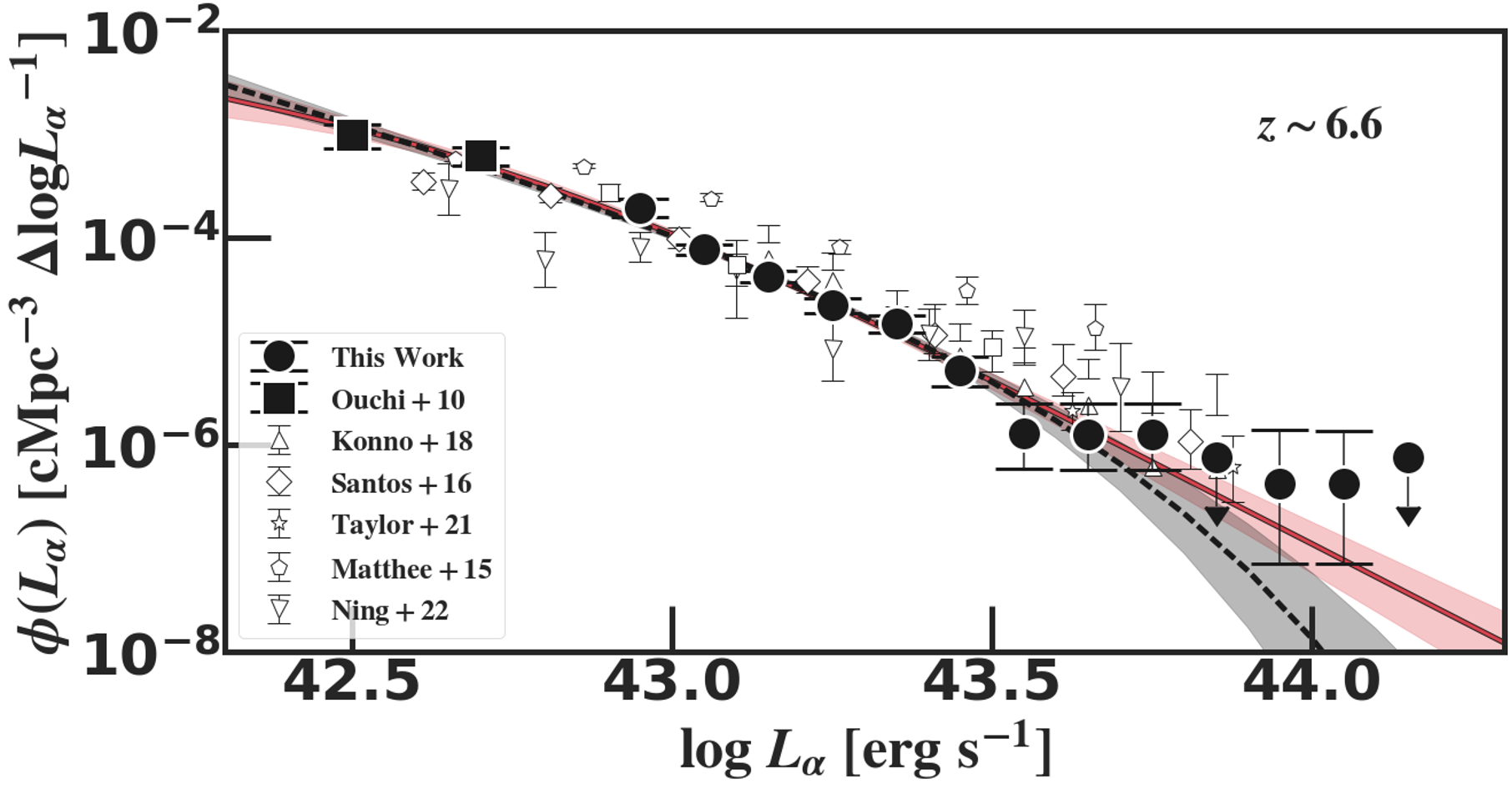}
    \caption{The fitting result for Ly$\alpha$ LF at $z=6.6$. The circle, squares, upward triangles, diamonds, stars, pentagons, and downward triangles with bars represent the number densities and corresponding errorbars from this work, \cite{Ouchi10}, \cite{Konno18}, \cite{Santos16}, \cite{Taylor21}, \cite{Matthee15}, and \cite{Ning22}, respectively. Filled (open) marks are \Add{(not)} used to fit the Ly$\alpha$ LF functions. The other symbols represent the same as in the Figure \ref{fit_lf_z4p9}.}
    \label{fit_lf_z6p6}
\end{figure*}

\begin{figure*}[htbp]
    \includegraphics[width=\linewidth]{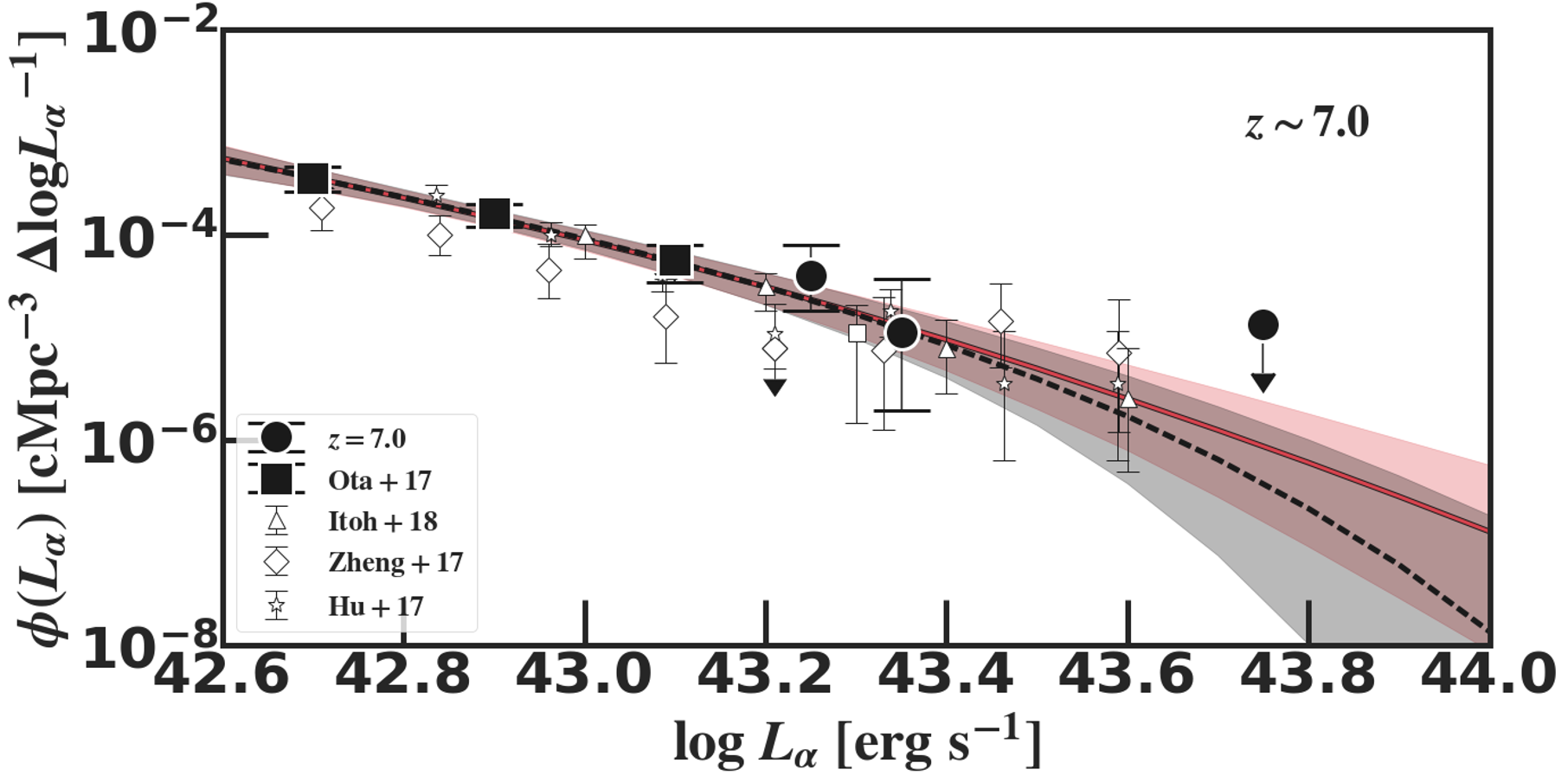}
    \caption{The fitting result for Ly$\alpha$ LF at $z=7.0$. The circle, squares, upward triangles, diamonds, and stars with bars represent the number densities and corresponding errorbars from this work, \cite{Ota17}, \cite{Itoh18}, \cite{Zheng17}, and \cite{Hu17}, respectively. Filled (open) marks are \Add{(not)} used to fit the Ly$\alpha$ LF functions. The other symbols represent the same as in the Figure \ref{fit_lf_z4p9}.}
    \label{fit_lf_z7p0}
\end{figure*}

\begin{figure*}[htbp]
    \includegraphics[width=\linewidth]{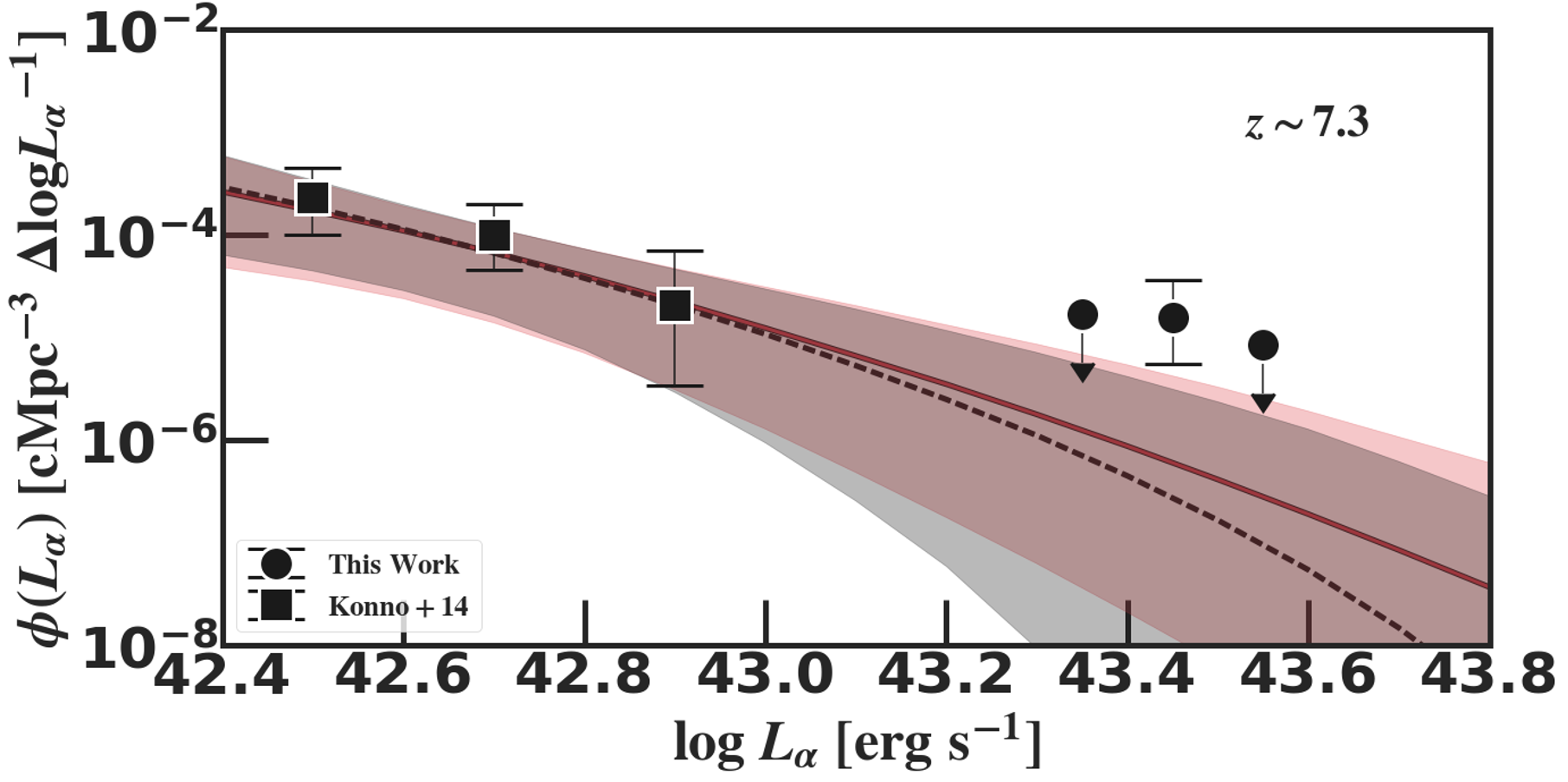}
    \caption{The fitting result for Ly$\alpha$ LF at $z=7.3$. The circles and squares with bars represent the number densities and corresponding errorbars from this work and \cite{Konno14}, respectively. Filled (open) marks are \Add{(not)} used to fit the Ly$\alpha$ LF functions. The other symbols represent the same as in the Figure \ref{fit_lf_z4p9}.}
    \label{fit_lf_z7p3}
\end{figure*}

In Figure\Add{s} \ref{fit_lf_z2p2} to \ref{fit_lf_z7p3}, we present the observed Ly$\alpha$ LFs together with fitted functions and other literature values not used in the fitting at $z=2.2$ to 7.3.
The data points used for the fitting is shown in the filled symbols, whereas the others are shown in the open symbols. While our fitted Ly$\alpha$ functions are mainly consistent with the observed values in this work and those from the literature, there are \Add{few notable} deviations between the observed number densities and the fitted Ly$\alpha$ luminosity functions. For example, the fitted results of Ly$\alpha$ LFs at $z=5.7$ do not agree with recent measurements using the Multi Unit Spectroscopic Explorer \citep[MUSE;][]{Bacon10} data \Add{by} \cite{Thai23} at $\log L_\alpha<42.0$. Although the exact reason for the discrepancy is unknown, the treatment of the detection completeness could cause the over/under correction of the number density at the faint end of luminosity function. Especially, for both \cite{Drake17} and \cite{Thai23}, the detection completeness values of the objects below $\log L_\alpha=42.0$ are smaller than 25\%, leading to the correction factor up to a few tens. Recovering the intrinsic LFs from such a low completeness regime could introduce systematics into the estimation, depending on how the detection completeness is calculated. \Add{Such a systematics could be a reason for the discrepancies we see between the literature values at the faint-end regimes.} We also see overabundant bright LAEs at the $z=7.3$. Number density at $\log L_\alpha\sim43.5$ exceeds more than 1$\sigma$ from the fitting results. \cite{Kikuta23} describes the detection of 5 LAEs at $z=7.3$ as ``tentative" due to their diffused appearances and low signal to noise ratio ($<5.4\sigma$) detection. However, if these overabundant source\Add{s} are real, then the bright-end excess could be caused by the contributions of Ly$\alpha$ emitters residing in the large ionized bubbles \citep[e.g.,][]{Saxena23}. \Add{We do also see offsets in $z=2.2$ Ly$\alpha$ luminosity function from this work, \cite{Konno16}, and \cite{Zhang21}, especially at around the bright-end. Such an offsets could come from the difference in the sample selection, as this work, \cite{Konno16}, and \cite{Zhang21} adopt different color selection. Additionally, the field-to-field variance could potentially be larger for LAEs than ordinary star-forming galaxies as discussed in \cite{Ma24}. We do note that the offsets at the bright end would not affect the following discussions in this work, and the intensive spectroscopic follow ups of $z=2.2$ LAE candidates by Subaru/PFS will enable the determination of the luminosity function shape through out wide luminosity range based on the consistent sample selection.}\par 
\subsection{Evolution of Ly$\alpha$ Luminosity Function}
\begin{figure*}[htbp]
    \includegraphics[width=\linewidth]{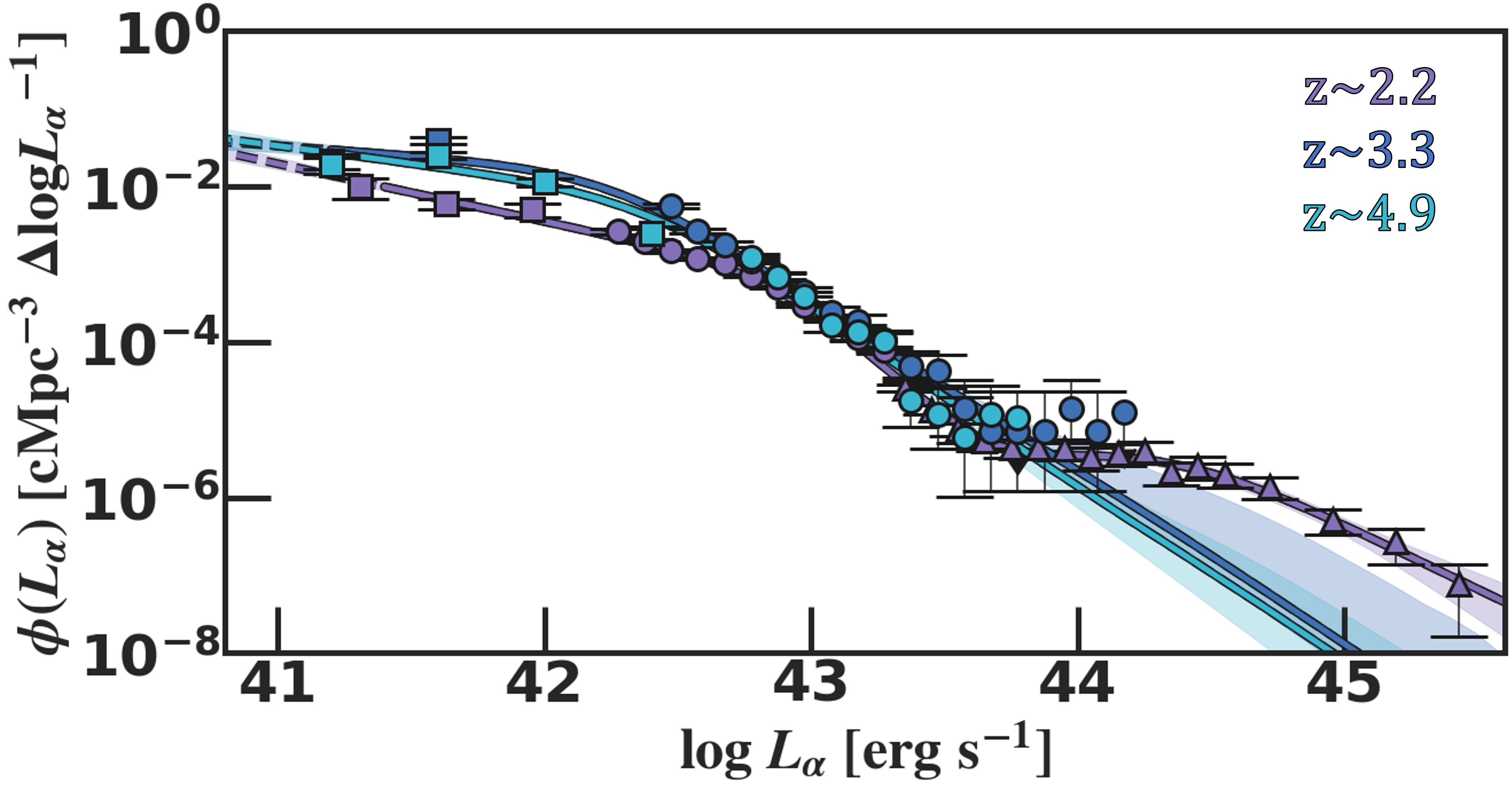}
    \caption{Evolution of Ly$\alpha$ luminosity function after the epoch of cosmic reionization. The Ly$\alpha$ luminosity function for $z$=2.2, 3.3, and 4.9 are shown in the purple, blue, and pale blue, respectively. The solid (dashed) lines represent the median of the fitted DPL + DPL (Schechter + DPL) functions in the luminosity range (not) used in the fitting. The shades represent 68-th percentile range for the fitted functions at the redshift of corresponding color. The symbols in each color represent the same as in the corresponding figures in Figure\Add{s} \ref{fit_lf_z2p2}, \ref{fit_lf_z3p3}, and \ref{fit_lf_z4p9}, respectively.}
\label{afterEoR_LF_fit}
\end{figure*}

\begin{figure*}[htbp]
    \includegraphics[width=\linewidth]{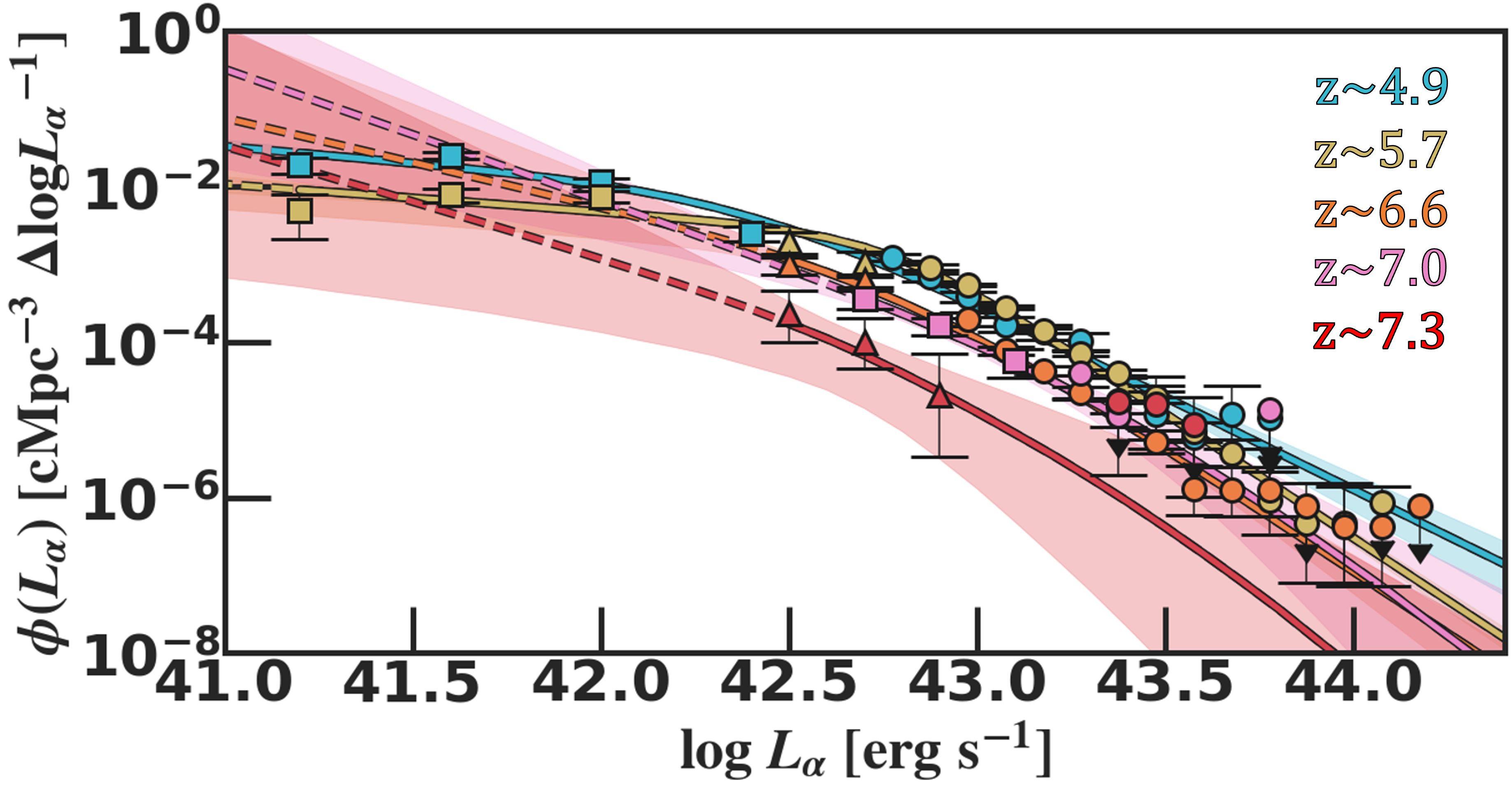}
    \caption{Evolution of Ly$\alpha$ luminosity function after the epoch of cosmic reionization. The Ly$\alpha$ luminosity function for $z$=4.9, 5.7, 6.6, 7.0, and 7.3 are shown in the pale-blue, yellow, orange, pink, and red, respectively. The solid (dashed) lines represent the median of the fitted DPL (\Add{Schechter}) functions in the luminosity range (not) used in the fitting. The shades represent 68-th percentile range for the fitted functions at the redshift of corresponding color. The symbols in each color represent the same as in the corresponding figures in Figure\Add{s} \ref{fit_lf_z4p9}, \ref{fit_lf_z5p7}, \ref{fit_lf_z6p6}, \ref{fit_lf_z7p0}, and \ref{fit_lf_z7p3}, respectively.}
    \label{EoR_LF_fit}
\end{figure*}

\begin{figure}[htbp]
    \includegraphics[width=\linewidth]{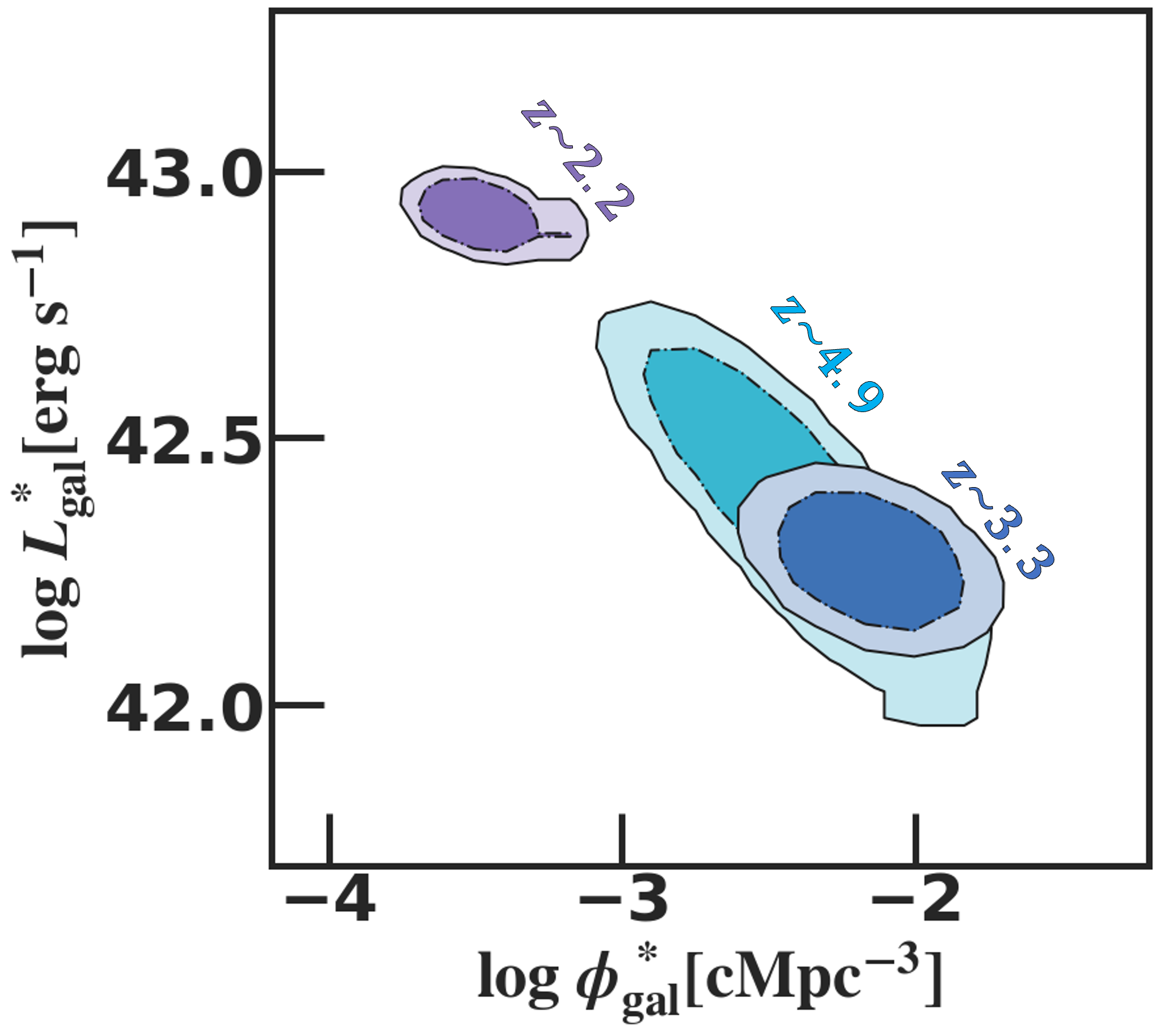}
    \caption{The 2-D marginalized PDFs for the $\log \phi^{\star}_{\rm gal}$ and $\log L^{\star}_{\rm gal}$ when \Add{DPL} (DPL) function is fitted for galaxy (AGN) component. \Add{The purple, blue, and pale-blue contours represent the 2-D margnialized PDFs for Ly$\alpha$ luminosity function at $z=2.2$, 3.3, and 4.9, respectively.} The dark and light component for each contour represents 68-th and 90-th percentile ranges, respectively.}
    \label{Sch_params_AoR}
\end{figure}

\begin{figure}[htbp]
    \includegraphics[width=\linewidth]{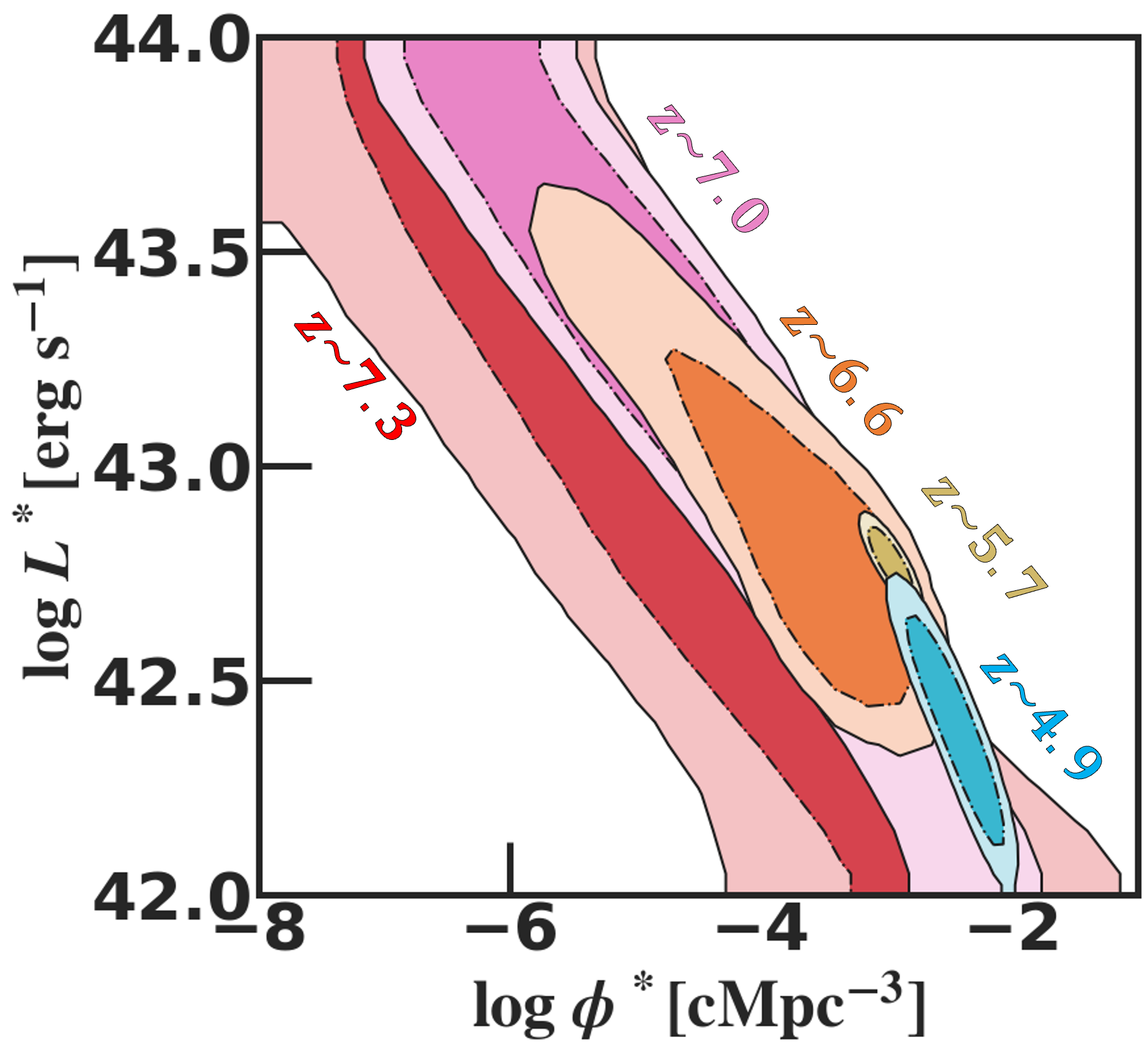}
    \caption{The 2-D marginalized PDFs for the $\log \phi^{\star}$ and $\log \phi^{\star}$ when Schechter function is fitted. The pale-blue, yellow, orange, pink, and red contours represent the 2-D marginalized PDFs for Ly$\alpha$ luminosity function at $z=4.9$, 5.7, 6.6, 7.0, and 7.3, respectively. The dark and light component for each contour represents 68-th and 90-th percentile ranges, respectively.}
    \label{Sch_params_EoR}
\end{figure}

We also show the 2-D marginalized PDFs for $\log \phi^\star_{\rm gal}$ and $\log L^\star_{\rm gal}$ at $z=2.2$, 3.3, and 4.9 for Schechter + DPL fitting functions in Figure \ref{Sch_params_AoR}. We show the 2-D marginalized PDFs for $\log \phi^\star$ and $\log L^\star$ at $z=4.9$, 5.7, 6.6, 7.0, and 7.3 in Figure \ref{Sch_params_EoR}. We summarize the fitted luminosity functions from $z=2.2-4.9$ ($z=4.9-7.3$) for DPL + DPL (DPL) fitting functions in Figure \ref{afterEoR_LF_fit} (\ref{EoR_LF_fit}). We see no significant evolution in the luminosity function at the luminosity range of $\log L_\alpha/{\rm erg~s^{-1}}=43.0-44.0$ through redshift \Add{$z=2.2$ to $5.7$. The Ly$\alpha$ luminosity functions start to slight decrease from $z=5.7$ to $7.0$.} However, as discussed in \cite{Konno14}, we confirm the decrease in the number density from $z=7.0$ to 7.3 by $>1\sigma$ significance around $\log L_\alpha/{\rm erg~s^{-1}}= 42.5-43.0$. \par
\begin{figure}[htbp]
    \includegraphics[width=\linewidth]{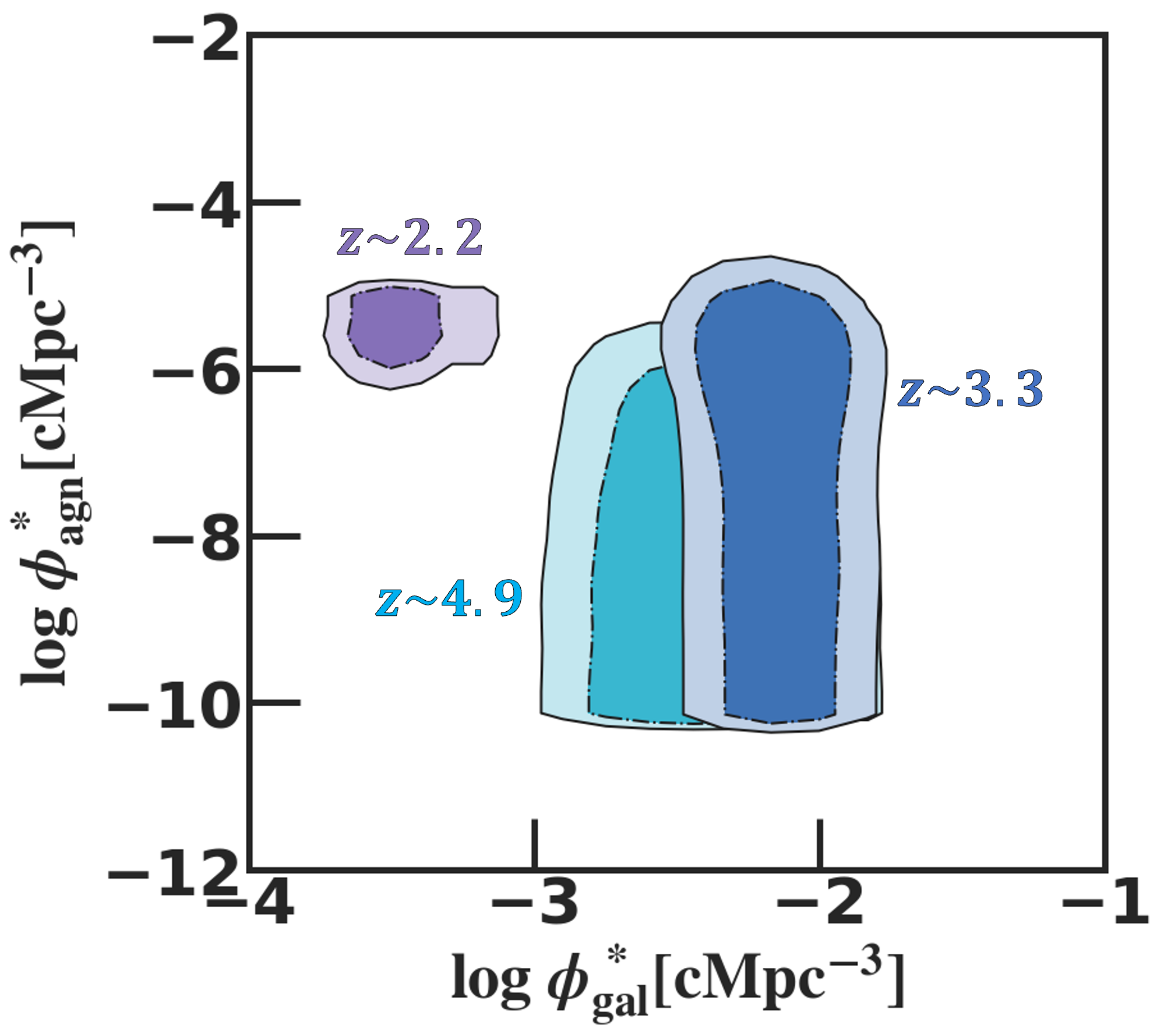}
    \caption{The 2-D marginalized PDFs for the $\log \phi^{\star}_{\rm gal}$ and $\log \phi^{\star}_{\rm agn}$ when \Add{DPL} functions are fitted for galaxy and AGN components. The purple, blue, and pale-blue contours represent the 2-D marginalized PDFs for Ly$\alpha$ luminosity function at $z=2.2$, 3.3, and 4.9, respectively. The dark and light component for each contour represents 68-th and 90-th percentile ranges, respectively. \Add{Note that PDfs for $z=3.3$ and 4.9 reach the lower bound values for $\log \phi^\star_{\rm agn}$ prior (i.e., $\log \phi^\star_{\rm agn}/{\rm {cMpc}^{-3}}=10^{-10}$).}}
    \label{phi_agn_gal}
\end{figure}

In Figure \ref{phi_agn_gal}, also plot 2D marginalized PDFs for $\log \phi^\star_{\rm gal}$ and $\log \phi^\star_{\rm agn}$ for the case of \Add{DPL} + DPL fitting functions. \Add{We confirm that the AGN components start to dominate the Ly$\alpha$ luminosity function (i.e., $\phi_{\rm agn}(L)>\phi_{\rm gal}$) above $\log L_\alpha/{\rm erg~s^{-1}}= 43.7$ for $z=2.2$ LAE sample. The AGN component fraction would be higher if we adopt the number density from this work or \cite{Konno16} instead of \cite{Zhang21} for the bright-end.} However, as $\log \phi^\star_{\rm agn}$ at $z=3.3$ and 4.9 are consistent with $\log \phi^\star_{\rm agn}/{\rm cMpc}^3 < -8.0$ within 68-th percentile range, we infer that AGN components at the $z=3.3$ and 4.9 are \Add{not significant} compared with galaxy component. Hereafter, we \Add{adopt} the single component \Add{fitting function (i.e., single DPL/Schechter function) for $z=4.9$ luminosity function in the following discussion.}


\begin{deluxetable*}{ccccccccccc}
\tablecolumns{11}
\tabletypesize{\scriptsize}
\tablecaption{Best-fit Parameters for Ly$\alpha$ LF at $z=2.2-4.9$
\label{table:lf_params_AoR}}
\tablehead{
\colhead{$z$} & \colhead{Fitting Function} & \colhead{$\log\phi^\star_{\rm gal}$} & \colhead{$\log L^\star_{\rm gal}$} & \colhead{$\alpha_{\rm gal}$} & \colhead{$\beta_{\rm gal}$} & \colhead{$\log\phi^\star_{\rm agn}$} & \colhead{$\log L^\star_{\rm agn}$} & \colhead{$\alpha_{\rm agn}$} & \colhead{$\beta_{\rm agn}$} & \colhead{$\log\rho^{\rm Ly\alpha}_{\rm obs, gal}$} \\
& & (${\rm cMpc}^{-3}$) & $({\rm erg~s^{-1}})$ & & & (${\rm cMpc}^{-3}$) & $({\rm erg~s^{-1}})$ & & & (${\rm erg~s^{-1}~{cMpc}^{-3}}$)\\ \hline
(1) & (2) & (3) & (4) & (5) & (6) & (7) & (8) & (9) & (10) & (11)
}
\startdata 
2.2 & Schechter + DPL & ${-3.16}_{-0.06}^{+0.04}$ & ${42.80}_{-0.03}^{+0.03}$ & ${-1.53}_{-0.06}^{+0.06}$ & \nodata & ${-5.92}_{-0.25}^{+0.30}$ & ${44.64}_{-0.43}^{+0.24}$ & ${-1.28}_{-0.22}^{+0.34}$ & ${-3.12}_{-0.92}^{+0.60}$ & ${39.38}_{-0.01}^{+0.01}$ \\
2.2 & DPL + DPL & ${-3.50}_{-0.06}^{+0.06}$ & ${42.92}_{-0.04}^{+0.03}$ & ${-1.74}_{-0.06}^{+0.06}$ & ${-4.27}_{-0.18}^{+0.15}$ & ${-5.58}_{-0.25}^{+0.11}$ & ${44.29}_{-0.23}^{+0.35}$ & ${-0.40}_{-0.61}^{+0.76}$ & ${-2.65}_{-0.75}^{+0.33}$ & ${39.40}_{-0.01}^{+0.01}$ \\
3.3 & Schechter + DPL & ${-2.13}_{-0.10}^{+0.08}$ & ${42.29}_{-0.06}^{+0.06}$ & ${-1.19}_{-0.13}^{+0.11}$ & \nodata & ${-4.57}_{-0.38}^{+0.13}$ & ${43.57}_{-0.06}^{+0.29}$ & ${-1.88}_{-0.13}^{+0.16}$ & ${-4.19}_{-1.54}^{+0.97}$ & ${39.40}_{-0.06}^{+0.05}$ \\
3.3 & DPL + DPL & ${-2.14}_{-0.10}^{+0.10}$ & ${42.27}_{-0.07}^{+0.07}$ & ${-1.23}_{-0.11}^{+0.11}$ & ${-3.30}_{-0.13}^{+0.12}$ & ${-7.23}_{-1.99}^{+1.59}$ & ${44.16}_{-0.49}^{+0.54}$ & ${-0.78}_{-1.16}^{+1.25}$ & ${-4.31}_{-2.46}^{+2.10}$ & ${39.66}_{-0.03}^{+0.02}$ \\
4.9 & Schechter + DPL & ${-2.72}_{-0.16}^{+0.14}$ & ${42.64}_{-0.08}^{+0.08}$ & ${-1.57}_{-0.09}^{+0.10}$ & \nodata & ${-6.06}_{-2.15}^{+0.82}$ & ${44.07}_{-0.44}^{+0.60}$ & ${-1.39}_{-0.70}^{+1.44}$ & ${-4.69}_{-2.20}^{+1.96}$ & ${39.53}_{-0.04}^{+0.03}$ \\
4.9 & DPL + DPL & ${-2.47}_{-0.28}^{+0.25}$ & ${42.41}_{-0.19}^{+0.19}$ & ${-1.44}_{-0.17}^{+0.20}$ & ${-3.41}_{-0.34}^{+0.23}$ & ${-7.99}_{-1.40}^{+1.64}$ & ${44.16}_{-0.46}^{+0.55}$ & ${-0.74}_{-1.39}^{+1.21}$ & ${-4.57}_{-2.37}^{+2.11}$ & ${39.57}_{-0.03}^{+0.03}$ \\
\enddata
\tablecomments{(1): Redshift of the sample. (2): Fitting function for Ly$\alpha$. Schechter function or double power-law (DPL).  (3),(7): Normalization factor for galaxy and AGN component, respectively. (4),(8): Characteristic Ly$\alpha$ luminosity for galaxy and AGN component, respectively. (5),(9): Faint-end slope for galaxy and AGN component, respectively. (6),(10): Bright-end slope (only for DPL) for galaxy and AGN component, respectively. (11): Ly$\alpha$ luminosity density obtained by integrating Ly$\alpha$ LF down to $\log L\Add{/{\rm erg~{s}^{-1}}}=42.5$ for galaxy component only.}
\end{deluxetable*}

\begin{deluxetable*}{ccccccc}
\tablecolumns{7}
\tabletypesize{\scriptsize}
\tablecaption{Best-fit Parameters for Ly$\alpha$ LF at $z=4.9-7.3$
\label{table:lf_params_EoR}}
\tablehead{
\colhead{$z$} & \colhead{Fitting Function} & \colhead{$\log\phi^\star$} & \colhead{$\log L^\star$} & \colhead{$\alpha$} & \colhead{$\beta$} & \colhead{$\log\rho^{\rm Ly\alpha}_{\rm obs}$} \\
& & (${\rm cMpc}^{-3}$) & $({\rm erg~s^{-1}})$ & & & (${\rm erg~s^{-1}~{cMpc}^{-3}}$)\\ \hline
(1) & (2) & (3) & (4) & (5) & (6) & (7)
}
\startdata 
4.9 & Schechter & ${-2.79}_{-0.12}^{+0.11}$ & ${42.69}_{-0.06}^{+0.06}$ & ${-1.61}_{-0.07}^{+0.07}$ & \nodata & ${39.55}_{-0.03}^{+0.02}$ \\
4.9 & DPL & ${-2.45}_{-0.27}^{+0.25}$ & ${42.40}_{-0.18}^{+0.18}$ & ${-1.43}_{-0.16}^{+0.20}$ & ${-3.38}_{-0.31}^{+0.22}$ & ${39.57}_{-0.03}^{+0.02}$ \\
5.7 & Schechter & ${-2.91}_{-0.05}^{+0.05}$ & ${42.75}_{-0.03}^{+0.03}$ & ${-1.39}_{-0.06}^{+0.06}$ & \nodata & ${39.56}_{-0.01}^{+0.01}$ \\
5.7 & DPL & ${-2.97}_{-0.11}^{+0.11}$ & ${42.78}_{-0.05}^{+0.05}$ & ${-1.35}_{-0.09}^{+0.10}$ & ${-4.20}_{-0.13}^{+0.12}$ & ${39.64}_{-0.03}^{+0.03}$ \\
6.6 & Schechter & ${-4.62}_{-0.74}^{+0.50}$ & ${43.28}_{-0.18}^{+0.26}$ & ${-2.82}_{-0.28}^{+0.32}$ & \nodata & ${39.13}_{-0.04}^{+0.04}$ \\
6.6 & DPL & ${-3.57}_{-1.09}^{+0.38}$ & ${42.81}_{-0.18}^{+0.43}$ & ${-2.13}_{-0.78}^{+0.78}$ & ${-4.17}_{-0.59}^{+0.30}$ & ${39.15}_{-0.04}^{+0.04}$ \\
7.0 & Schechter & ${-4.49}_{-1.27}^{+0.85}$ & ${43.23}_{-0.39}^{+0.49}$ & ${-2.49}_{-0.58}^{+0.82}$ & \nodata & ${39.04}_{-0.09}^{+0.07}$ \\
7.0 & DPL & ${-4.65}_{-1.18}^{+1.28}$ & ${43.24}_{-0.61}^{+0.47}$ & ${-2.68}_{-0.52}^{+0.84}$ & ${-4.51}_{-2.19}^{+1.11}$ & ${39.05}_{-0.09}^{+0.08}$ \\
7.3 & Schechter & ${-5.00}_{-1.51}^{+1.40}$ & ${43.03}_{-0.61}^{+0.62}$ & ${-2.56}_{-0.90}^{+0.93}$ & \nodata & ${38.29}_{-0.57}^{+0.27}$ \\
7.3 & DPL & ${-5.03}_{-1.44}^{+1.42}$ & ${42.98}_{-0.64}^{+0.64}$ & ${-2.51}_{-0.89}^{+0.92}$ & ${-5.02}_{-1.97}^{+1.64}$ & ${38.31}_{-0.63}^{+0.28}$ \\
\hline
\enddata
\tablecomments{(1): Redshift of the sample. (2): Fitting function for Ly$\alpha$. Schechter function or double power-law (DPL).  (3): Normalization factor. (4): Characteristic Ly$\alpha$ luminosity. (5): Faint-end slope. (6): Bright-end slope (only for DPL). (7): Ly$\alpha$ luminosity density obtained by integrating Ly$\alpha$ LF down to $\log L/\Add{{\rm erg~{s}^{-1}}}=42.4$.}
\end{deluxetable*}

\begin{deluxetable*}{c|ccccccc}
\tablecolumns{8}
\tabletypesize{\scriptsize}
\tablecaption{Ly$\alpha$ LF at $z=2.2-7.3$ from This Work
\label{table:num_den}}
\tablehead{
\multicolumn{1}{c}{$\log L_\alpha$} & 
\multicolumn{7}{c}{$\log \phi(L_\alpha)$} \\
\colhead{} & \colhead{$z\sim2.2$} & \colhead{$z~\sim3.3$} & \colhead{$z\sim4.9$} & \colhead{$z\sim5.7$} & \colhead{$z\sim6.6$} & \colhead{$z\sim7.0$} & \colhead{$z\sim7.3$} \\ \hline
\multicolumn{1}{c}{($\mathrm{erg}~\mathrm{s}^{-1}$)} & 
\multicolumn{7}{c}{$(\mathrm{cMpc}^{-3}~\Delta \log L_{\alpha}^{-1})$}
}
\startdata 
42.25& $-2.59_{-0.03}^{+0.03}$& \nodata& \nodata& \nodata& \nodata& \nodata& \nodata\\
42.35& $-2.71_{-0.02}^{+0.02}$& \nodata& \nodata& \nodata& \nodata& \nodata& \nodata\\
42.45& $-2.84_{-0.02}^{+0.02}$& $-2.25_{-0.02}^{+0.02}$& \nodata& \nodata& \nodata& \nodata& \nodata\\
42.55& $-2.93_{-0.01}^{+0.01}$& $-2.58_{-0.03}^{+0.03}$& \nodata& \nodata& \nodata& \nodata& \nodata\\
42.65& $-2.99_{-0.01}^{+0.01}$& $-2.75_{-0.03}^{+0.03}$& \nodata& \nodata& \nodata& \nodata& \nodata\\
42.75& $-3.15_{-0.02}^{+0.02}$& $-2.96_{-0.04}^{+0.04}$& $-2.91_{-0.05}^{+0.05}$& \nodata& \nodata& \nodata& \nodata\\
42.85& $-3.30_{-0.02}^{+0.02}$& $-3.15_{-0.05}^{+0.05}$& $-3.17_{-0.05}^{+0.05}$& $-3.05_{-0.04}^{+0.04}$& \nodata& \nodata& \nodata\\
42.95& $-3.53_{-0.02}^{+0.02}$& $-3.36_{-0.06}^{+0.06}$& $-3.42_{-0.06}^{+0.06}$& $-3.27_{-0.02}^{+0.02}$& $-3.71_{-0.09}^{+0.09}$& \nodata& \nodata\\
43.05& $-3.72_{-0.03}^{+0.03}$& $-3.62_{-0.08}^{+0.08}$& $-3.78_{-0.09}^{+0.09}$& $-3.56_{-0.02}^{+0.02}$& $-4.11_{-0.05}^{+0.05}$& \nodata& \nodata\\
43.15& $-3.95_{-0.04}^{+0.04}$& $-3.74_{-0.09}^{+0.09}$& $-3.88_{-0.10}^{+0.10}$& $-3.85_{-0.03}^{+0.03}$& $-4.37_{-0.06}^{+0.06}$& \nodata& \nodata\\
43.25& $-4.11_{-0.04}^{+0.04}$& $-3.98_{-0.13}^{+0.12}$& $-4.00_{-0.12}^{+0.12}$& $-4.15_{-0.04}^{+0.04}$& $-4.65_{-0.07}^{+0.07}$& $-4.40_{-0.34}^{+0.29}$& \nodata\\
43.35& $-4.26_{-0.05}^{+0.05}$& $-4.31_{-0.20}^{+0.19}$& $-4.75_{-0.34}^{+0.29}$& $-4.41_{-0.05}^{+0.05}$& $-4.83_{-0.08}^{+0.08}$& $-4.95_{-0.77}^{+0.52}$& $<-4.77$\\
43.45& $-4.42_{-0.06}^{+0.06}$& $-4.38_{-0.22}^{+0.20}$& $-4.93_{-0.45}^{+0.36}$& $-4.72_{-0.07}^{+0.07}$& $-5.28_{-0.15}^{+0.14}$& $<-4.87$& $-4.81_{-0.45}^{+0.36}$\\
43.55& $-4.60_{-0.08}^{+0.08}$& $-4.86_{-0.45}^{+0.36}$& $-5.23_{-0.77}^{+0.52}$& $-5.15_{-0.13}^{+0.12}$& $-5.89_{-0.34}^{+0.29}$& $<-4.87$& $<-5.07$\\
43.65& $-4.64_{-0.08}^{+0.08}$& $-5.16_{-0.77}^{+0.52}$& $-4.93_{-0.45}^{+0.36}$& $-5.43_{-0.18}^{+0.17}$& $-5.90_{-0.34}^{+0.29}$& $<-4.87$& $<-5.07$\\
43.75& $-4.81_{-0.10}^{+0.10}$& $-5.16_{-0.77}^{+0.52}$& $<-4.98$& $-6.03_{-0.45}^{+0.36}$& $-5.90_{-0.34}^{+0.29}$& $<-4.87$& $<-5.07$\\
43.85& $-4.81_{-0.10}^{+0.10}$& $-5.16_{-0.77}^{+0.52}$& $<-4.98$& $-6.33_{-0.77}^{+0.52}$& $<-6.12$& $<-4.87$& $<-5.07$\\
43.95& $-4.78_{-0.10}^{+0.10}$& $-4.86_{-0.45}^{+0.36}$& $<-4.98$& $-6.33_{-0.77}^{+0.52}$& $-6.38_{-0.77}^{+0.52}$& $<-4.87$& $<-5.07$\\
44.05& $-4.90_{-0.12}^{+0.11}$& $-5.16_{-0.77}^{+0.52}$& $<-4.98$& $<-6.07$& $-6.38_{-0.77}^{+0.52}$& $<-4.87$& $<-5.07$\\
44.15& $-5.04_{-0.14}^{+0.13}$& $<-4.90$& $<-4.98$& $<-6.07$& $<-6.12$& $<-4.87$& $<-5.07$\\
44.25& $-5.16_{-0.16}^{+0.15}$& $<-4.90$& $<-4.98$& $<-6.07$& $<-6.12$& $<-4.87$& $<-5.07$\\
44.35& $-5.01_{-0.13}^{+0.13}$& $<-4.90$& $<-4.98$& $<-6.07$& $<-6.12$& $<-4.87$& $<-5.07$\\
44.45& $-5.20_{-0.17}^{+0.16}$& $<-4.90$& $<-4.98$& $<-6.07$& $<-6.12$& $<-4.87$& $<-5.07$\\
44.55& $-5.38_{-0.22}^{+0.20}$& $<-4.90$& $<-4.98$& $<-6.07$& $<-6.12$& $<-4.87$& $<-5.07$\\
44.65& $-5.68_{-0.34}^{+0.29}$& $<-4.90$& $<-4.98$& $<-6.07$& $<-6.12$& $<-4.87$& $<-5.07$\\
44.75& $-5.55_{-0.28}^{+0.25}$& $<-4.90$& $<-4.98$& $<-6.07$& $<-6.12$& $<-4.87$& $<-5.07$\\
44.85& $-6.16_{-0.77}^{+0.52}$& $<-4.90$& $<-4.98$& $<-6.07$& $<-6.12$& $<-4.87$& $<-5.07$\\
44.95& $-6.16_{-0.77}^{+0.52}$& $<-4.90$& $<-4.98$& $<-6.07$& $<-6.12$& $<-4.87$& $<-5.07$\\
\enddata
\tablecomments{The number density of Ly$\alpha$ emitters $\log \phi(L_\alpha)$ at each Ly$\alpha$ luminosity for $z=2.2$, 3.3, 4.9, 5.7, 6.6, 7.0, and 7.3. The luminosity bin our our $Ly\alpha$ LFs is $\Delta \log~L_\alpha=0.1$. The number densities are corrected for the detection completeness. The number density below the limiting luminosity are not shown.}
\end{deluxetable*}

\section{Clustering Analysis} \label{sec:cluster}

\subsection{Angular Correlation Function}\label{acf_expl}
Cosmic reionization also leaves its trace on the clustering properties of observed LAEs. \Add{In inside-out cosmic reionization scenarios, LAE clustering generally strengthens as the $x_{\rm \HI}$ increases because LAEs residing inside the ionized bubbles are selectively observed due to less Ly$\alpha$ obscuration \citep[e.g.,][]{Furlanetto06,McQuinn07,Ouchi18}.} In this section, we estimate \Add{the spatial auto-correlation} function of LAEs at different redshifts to investigate the influence of cosmic reionization on the clustering properties and estimate $x_{\rm \HI}$. \par
Because we do not have precise redshift information for each object in the photometric sample, we calculate the \Add{angular correlation function} (ACF) instead of the three-dimensional spatial correlation function. We calculate the ACF in the same manner as \cite{Ouchi18}.
To estimate ACF for angular separation of $\theta$ ($\omega(\theta)$), we adopt the estimator presented in \cite{1993ApJ...412...64L},
\begin{equation}
  \omega_{\rm obs}(\theta)=\frac{DD(\theta)-2DR(\theta)-RR(\theta)}{RR(\theta)},
  \label{omega_obs}
\end{equation}
where $DD(\theta)$, $RR(\theta)$, and $DR(\theta)$ are the number counts
of the galaxy-galaxy, random-random, and galaxy-random pairs with the projected spatial separation of the angular scale $\theta$, respectively. For the random sample, we use a random catalog provided by the HSC DR4 database. This random catalog has the position of points randomly extracted from the same geometry as that of the LAE survey field with a number density of 100 ${\rm arcmin}^{-2}$. \Add{To minimize the systematics introduced from the spatially inhomogeneous detection completeness across the survey area, we apply a luminosity cut to the galaxy sample when we calculate ACFs. We only use galaxies with $NB$ magnitude brighter than 25 mag, which corresponds to the Ly$\alpha$ luminosity around the characteristic limiting luminosity at $z=4.9-6.6$. We summarize the sample size and the luminosity cut we adopt at each redshift for the ACF calculations in Table \ref{table:clus}.} We correct the estimator of Eq. \ref{omega_obs} for the systematics introduce\Add{d} by the limited survey field and the sample size of galaxies in the following manner:
\begin{equation}
    \omega(\theta)=\omega_{\rm obs}(\theta) + \mathrm{IC} + 1/N_\mathrm{LAE},
\end{equation}
where IC is the integral constraint and $N_\mathrm{LAE}$ is the sample size of galaxies. Following \cite{1977ApJ...217..385G}, we calculate the integral constraint as follows:
\begin{equation}
    \mathrm{IC} = \sum_{i}{RR(\theta_i)}\omega_{\rm mod}(\theta_i)/\sum_i{RR(\theta_i)},
\end{equation}
where $\omega_{\rm mod}$ represents the model prediction for ACF. \par
For a spatial correlation function model ($\xi$), we assume a simple power-law function,
\begin{equation}
  \xi_{\rm mod}(r) ={\left(\frac{r}{r_0}\right)}^{-\gamma},
  \label{omega_pl}
\end{equation}
where $r$, $r_0$, and $\gamma$ are the spatial separation between two galaxies,
the characteristic correlation length, and the power-law index, respectively.
We use the numerical method presented in \cite{2007A&A...473..711S} to convert 3-D spatial correlation functions to ACF. After substituting equation \ref{omega_pl} into Simon's equation, we end up having the model function in the form of
\begin{equation}
\begin{aligned}
  \omega_{\rm mod}(\theta)=\int_0^\infty dr_1 \int_0^\infty &dr_2 \xi(|\vec{r_1} - \vec{r_2}|)p(r_1)p(r_2) \\
  &\times{(r_1^2+r_2^2-2r_1r_2\cos\theta)}^{-\frac{\gamma}{2}},
\end{aligned}
\label{simon}
\end{equation}
where $r_1$ ($r_2$) and $p(r)$ are the
comoving radial distance to the first (second) galaxy of the pair and the marginalized probability of finding a galaxy at the radial comoving distance $r$, respectively.
We adopt the normalized narrowband transmission curve as $p(r)$.\par 

We estimate the unbiased covariance matrix of the ACF using Jackknife resampling. In our Jackknife estimation, we divide the survey fields into sub-fields with the area of $\sim1000^2~{\rm arcsec}^2$ each to create Jackknife samples. We remove one sub-sample each time and calculate the ACF with the rest of other sub-samples (i.e., Jackknife sample), then compute covariance matrix $C$ as
\begin{equation}
C_{i,j}=\frac{N_{\rm Jack}-1}{N_{\rm Jack}}\sum_{l=1}^{N_{\rm Jack}}[\omega^l(\theta_i)-\overline{\omega}(\theta_i)][\omega^l(\theta_j)-\overline{\omega}(\theta_j)],
\label{jackknife}
\end{equation}
where $N_{\rm Jack}$ is the total number of the Jackknife sample, $\omega^l$ is the estimated correlation function from $l-$th calculation, and $\overline{\omega}$ is the average ACF. We use the inverse covariance matrix to calculate the chi-square between observed and model $\omega$. When we calculate the inverse covariance matrix, we correct for the bias introduced by the noise by applying a factor given by \cite{2007A&A...464..399H}. We calculate the chi-square value for different ($r_0$, $\gamma$) sets. We explore $r_0$ values from 0.0 to $9.99*0.7~h^{-1}~{\rm Mpc}$ with a stepsize of $0.7~h^{-1}~{\rm Mpc}$. We adopt 0.7 step size for the convenience of comparing the results with the literature values derived with an assumption that hubble constants is at 100 km/s/Mpc. Because previous studies found $\gamma\sim1.8$ for ACFs of LAEs \citep{2007ApJ...671..278G,2007ApJ...668...15K,Ouchi10},
we explore $\gamma$ values around 1.8, specifically from 0.50 to 2.49 with a stepsize of 0.01. We define the best-fit value for ($r_0$, $\gamma$) as those corresponding to the minimum chi-square ($\chi^2_{\rm min}$). The errorbars are given based on the 1 sigma chi-square value ranges (i.e., $\chi^2\leq\chi^2_{\rm min} + 1$). \Add{We fit in the range from $1.4<\log \theta/{\rm arcsec}<2.8$ to avoid the systematics introduced by the one-halo term at small scales and the limited survey area (i.e., Jackknife sub-field area $\sim1000^2~{\rm arcsec}^2$) in the large scale.}\par

\begin{figure*}[htbp]
    \centering
    \includegraphics[width=0.9\linewidth]{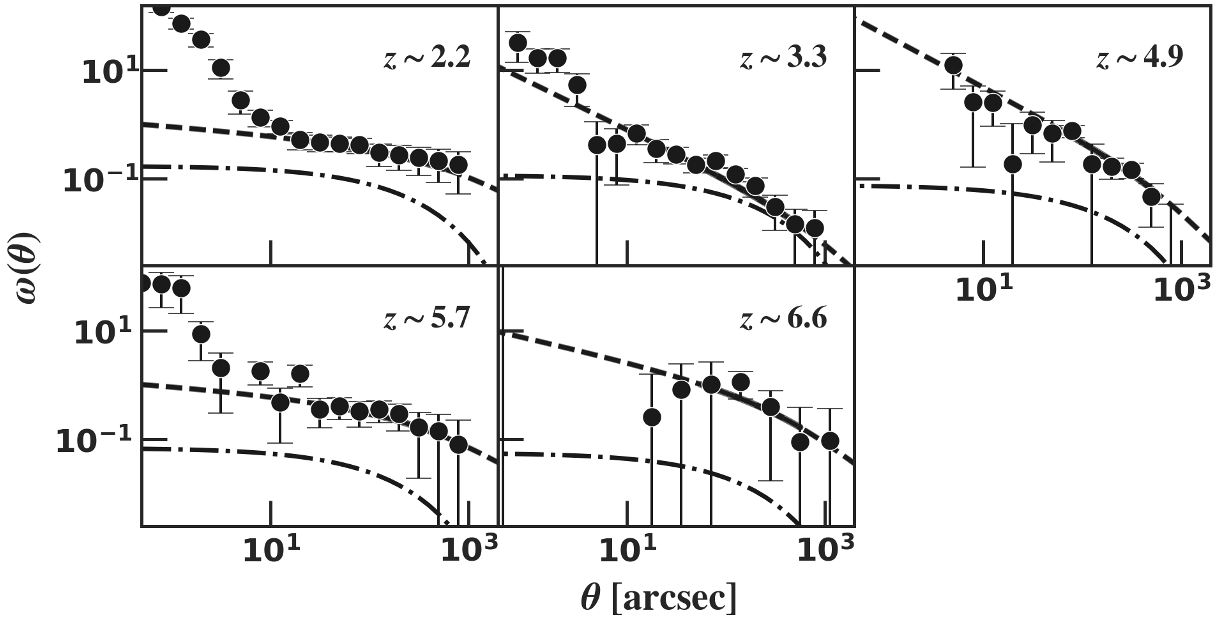}
    \caption{Two point angular auto-correlation function $\omega(\theta)$ of LAEs at $z=2.2$, 3.3, 4.9, 5.7, and 6.6. Black circles with bars represent the observed correlation functions and errors. The solid and dashed lines \Add{represent} the best-fit power-law model in and out of the the fitting range used, respectively. The dotted-dashed line represent the correlation function of the dark matter density fluctuation $\omega_{\rm DM}(\theta)$ at each redshift.}
\label{acf}
\end{figure*}

In Figure \ref{acf}, we present the observed and best-fit model values for ACFs at $z$=2.2, 3.3, 4.9, 5.7, and 6.6. We adopt the square roots of the diagonal elements of the covariance matrix as the uncertainties of the observed ACF. \par

\subsection{Correlation Length and Bias}
In Table \ref{table:clus}, we present the best-fit values together with the properties of data used to derive ACFs at different redshifts.\par
From the derived ACFs, we calculate the galaxy-dark matter bias ($b_g$), which is defined as follows:
\begin{equation}
    b_g^{2}(\theta) \equiv \omega(\theta)/\omega_{\rm DM}(\theta).
\end{equation}
Here, $\omega_{\rm DM}$ is the angular correlation function of the dark matter density fluctuation. $\omega_{\rm DM}$ can be calculated from spatial correlation function of dark matter density fluctuation $\xi_{\rm DM}(r)$ in the same manner as $\omega(\theta)$ from $\xi(r)$ in the equation \ref{simon}.
We calculate $\xi_{\rm DM}(r)$ at the redshift $z$ by
\begin{equation}
\xi_{\rm DM}(r,z) = \int \frac{k^2dk}{2\pi^2}\frac{\sin(kr)}{kr}P(k,z),
\end{equation}
where $P(k,z)$ is a linear perturbation spectrum of dark matter. Here, we adopt the Harrison-Zeldovich spectrum \citep{PhysRevD.1.2726} as $P(k,z)$. 
In Table \ref{table:clus}, we report the average bias $b_{\rm avg}$ by taking error weighted average of $b_{g}(\theta)$ over the angular range used in the fitting.

\begin{deluxetable*}{ccccccccc}
\tablecolumns{11}
\tabletypesize{\scriptsize}
\tablecaption{Clustering measurements for LAEs
\label{table:clus}}
\tablehead{
\colhead{$z$} & \colhead{$\log L_{\alpha, \rm th}$} & \colhead{$N$} & \colhead{IC} & \colhead{$r_0$} & \colhead{$b_{g,{\rm avg}}$} &  \colhead{$\gamma$} & \colhead{$\chi^2$} & \colhead{d.o.f.}\\
& ($\rm erg~s^{-1}$) & & ($10^{-2}$) & ($h^{-1}_{70}~{\rm Mpc}^{-3}$)  & \\ \hline
(1) & (2) & (3) & (4) & (5) & (6) & (7) & (8) & (9) 
}
\startdata 
2.2 & 42.0 & 3403 & 6.98 & ${3.80}_{-0.40}^{+0.50}$ & $1.40\pm0.10$ & ${0.87}_{-0.18}^{+0.19}$  & 4.84 & 5 \\
3.3 & 42.3 & 598 & 0.43 & ${2.70}_{-0.40}^{+0.30}$ & $1.59\pm0.20$ & ${1.89}_{-0.20}^{+0.28}$ & 4.96 & 5 \\
4.9 & 42.6 & 149 & 1.74 & ${6.40}_{-0.70}^{+0.60}$ & $3.69\pm0.40$ & ${1.97}_{-0.16}^{+0.16}$ & 3.89 & 5 \\
5.7 & 42.7 & 1584 & 3.94 & ${4.00}_{-0.70}^{+0.60}$ & $3.36\pm0.28$ & ${1.02}_{-0.17}^{+0.17}$ & 5.70 & 5 \\
6.6 & 42.8 & 159 & 2.94 & ${8.00}_{-5.80}^{+1.90}$ & $5.87\pm0.99$ & ${1.40}_{-0.88}^{+0.58}$ & 2.35 & 3 \\
\enddata
\tablecomments{(1): Redshifts \Add{(2): The threshold Ly$\alpha$ luminosity for LAE sample used in the clustering analysis. The threshold Ly$\alpha$ luminosity at each redshift sample corresponds to the $NB$ magnitude of 25 mag.} (3): Number of galaxies in the sample. (4): Integration Constraint. (5): Best fit value of the average correlation length. (6): Best fit value of the average galaxy bias for the scale $\lesssim 10~h_{70}^{-1}~{\rm Mpc}$. 
(7): Best fit value of the 
(8): $\chi^2$ value. (9): Degree of freedom. }
\end{deluxetable*}

\section{Discussions} \label{sec:disc}
\subsection{Ly$\alpha$ Equivalent Width Distributions Throughout Redshifts} \label{z_EW}
\begin{figure}[htbp]
    \centering
    \includegraphics[width=\linewidth]{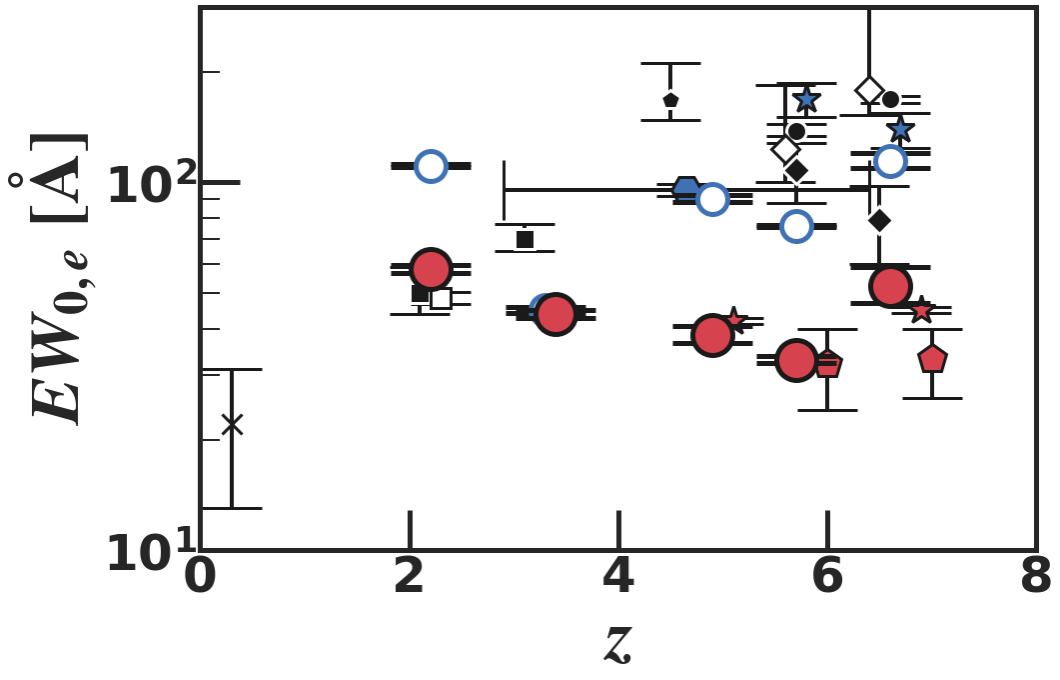}
    \caption{$e$-folding scale of Ly$\alpha$ equivalent width distribution at different redshifts. Our $EW_{0,e}$ constraints derived from the distribution fitting between 40 to 200 (1000) {\AA} data are shown in red (blue unfilled) circles. $EW_{0,e}$ values from \cite{Napolitano24} and \cite{Pentericci18} are shown in the red stars and pentagons, respectively. $EW_{0,e}$ values from \cite{Kerutt22} and \cite{Shibuya18a} are shown in the blue hexagon and stars, respectively. \Add{We also present \cite{Cowie10}, \cite{Ciardullo12}, \cite{Nilsson09}, \cite{Kashikawa11}, \cite{Hu10}, and \cite{Zheng14}, in the black x mark, white squares, black square, black diamonds, white diamonds, and black pentagon, respectively.}}
\label{EW_scale}
\end{figure}

\begin{deluxetable}{ccc}%c}
\tablecolumns{3}
\tabletypesize{\scriptsize}
\tablecaption{Ly$\alpha$ $EW_0$ Distribution's $e-$folding Scales
\label{table:EW0}}
\tablehead{
\colhead{$z$} & \colhead{Fitting Range} & \colhead{$EW_{0,e}$} \\
 & ({\AA}) & ({\AA}) \\
(1) & (2) & (3)
}
\startdata 
$2.2$ & 40-200 & ${58.2}_{-1.4}^{+1.4}$ \\
$2.2$ & 40-1000 & ${110.8}_{-0.9}^{+1.0}$ \\\hline
$3.3$ & 40-200 & ${43.9}_{-1.1}^{+1.1}$ \\
$3.3$ & 40-1000 & ${45.0}_{-0.8}^{+0.7}$ \\\hline
$4.9$ & 40-200 & ${38.5}_{-2.0}^{+2.2}$ \\
$4.9$ & 40-1000 & ${90.3}_{-1.7}^{+1.9}$ \\\hline
$5.7$ & 40-200 & ${32.9}_{-0.7}^{+0.8}$ \\
$5.7$ & 40-1000 & ${76.0}_{-0.6}^{+0.7}$ \\\hline
$6.6$ & 40-200 & ${52.3}_{-5.3}^{+6.5}$ \\
$6.6$ & 40-1000 & ${114.5}_{-4.9}^{+5.6}$
\enddata
\tablecomments{(1): Redshifts (2): Fitting range used to determine $EW_{0,e}$ values. (3): Inferred $EW_{0,e}$ values at each redshift.}
\end{deluxetable}
\Add{Multiple studies have found that the rest-frame Ly$\alpha$ equivalent width distributions can be well-explained in the exponential distribution $\propto \exp(-EW_{0}/EW_{0,e})$, where $EW_{0,e}$ is the $e-$folding scale. We fit the Ly$\alpha$ $EW_0$ distributions presented in Figure \ref{EW_dist} with an exponential distribution form using the least chi-square. We fit the distribution between $EW_0=40-200$ {\AA}. We do not include the $EW_0<40$ {\AA} data to avoid the impact from the color selection incompleteness. We report the best-fit $EW_{0,e}$ values obtained when fitting in this range for each redshift sample in the Table \ref{table:EW0}. We present our best-fit $EW_{0,e}$ at redshift $z=2.2$ to 6.6 in Figure \ref{EW_scale}. We also present literature $EW_{0,e}$ values from \cite{Napolitano24}, \cite{Pentericci18}, \cite{Cowie10}, \cite{Shibuya18a}, \cite{Ciardullo12}, \cite{Nilsson09}, \cite{Kashikawa11}, \cite{Hu10}, \cite{Zheng14}, and \cite{Kerutt22} in Figure \ref{EW_scale}. We can clearly see that our $EW_{0,e}$ values lie around 40 {\AA} across the redshift range, and it is consistent with some of the literature values \citep[e.g.,][]{Pentericci18} including the one from recent JWST measurements \citep{Napolitano24}. However, we do see distinctive populations of reported $EW_{0,e}$ values from multiple studies \citep{Shibuya14,Kerutt22} around $EW_{0,e}\approx$100 {\AA} above $z\sim3$ that deviate from our measurements. Using the same LAE sample, we performed the distribution fitting analysis with a higher upper limit of 1000 {\AA} and present the results in the Figure \ref{EW_scale}. Including high equivalent width objects (i.e., $EW_{0}>200$ {\AA}) systematically raises the $EW_{0,e}$ values up to around 100 {\AA}, more consistent with measurements from \cite{Shibuya18a} and \cite{Kerutt22} than those measured without including $EW_{0}>200$ {\AA} objects. We point out that the objects with $EW_0>$200 {\AA} represent only around or less than $5\%$ of the sample at all redshifts.}
\subsection{Redshift Evolution of Ly$\alpha$ Luminosity Functions} \label{z_lf}
In Figure\Add{s} \ref{Sch_params_AoR} and \ref{Sch_params_EoR}, we plot the posterior probability function for $\log \phi^\star$ and $\log L^\star$ for the \Add{luminosity} function at different redshifts. We do not see significant redshift evolution for $\log \phi^\star$ and $\log L^\star$ from $z=2.2$ to 5.7. However, the marginalized posterior distributions of $z=5.7$ and 7.3 do not overlap within both of the 68-th percentile range. This redshift evolution from $z=5.7$ to $z=7.3$ could be due to the increased in IGM attenuation strength at $z>7$ due to cosmic reionization as suggested in \cite{Konno14}. \par
\begin{figure*}[htbp]
    \centering
    \includegraphics[width=\linewidth]{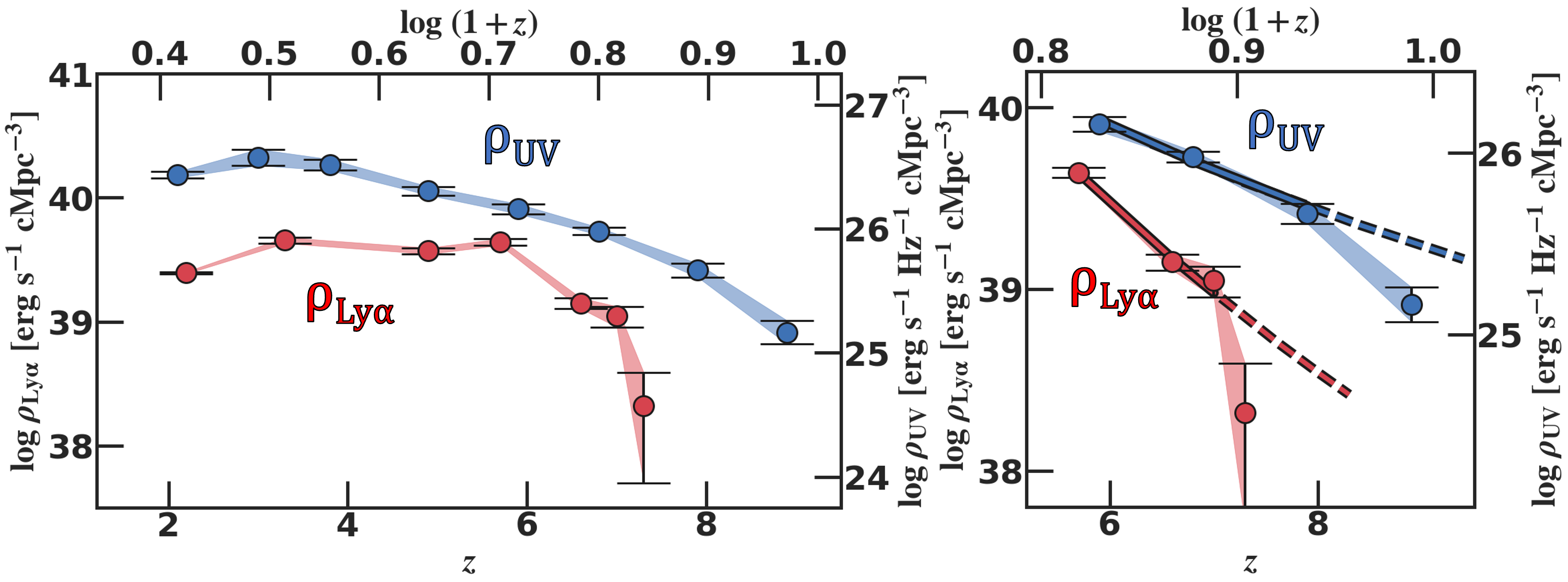}
    \caption{\Add{(Left) Redshift evolution of Ly$\alpha$ and dust-attenuated UV luminosity densities. The red circles and bars represent the Ly$\alpha$ luminosity densities and the uncertainties derived by integrating the fitted DPL function down to the Ly$\alpha$ luminosity of $L_{\rm cut}/{\rm erg~{s}^{-1}}=42.5$. The blue circles and bars represent the UV luminosity densities from \cite{Bouwens22}
 derived by integrating down to the UV luminosity of $M_{\rm UV}=-17.0$ mag. Red and blue shades corresponds to the uncertainty ranges for Ly$\alpha$ and UV luminosity densities across the redshifts, respectively. (Right) Zoom up view of the redshift evolution of Ly$\alpha$ and UV luminosity densities at the EoR. The solid red (blue) line represent the best-fit $\propto(1+z)^{\xi}$ function with $\xi$ as a free parameter fitted by the Ly$\alpha$ (UV) luminosity densities at $z=4.9-7$ ($z=6-8$). The dotted lines are the extrapolation of the fitted $\propto(1+z)^{\xi}$ functions.}}
    \label{lum_den_evolution}
\end{figure*}

 
To see the trend more clearly, we calculate Ly$\alpha$ luminosity density ($\rho_{\mathrm{Ly}\alpha}$) by integrating Ly$\alpha$ luminosity down to the cutoff luminosity ($L_{\rm cut}$) as follows:
\begin{equation}
\rho_{\mathrm{Ly}\alpha}=\int_{L_{\rm cut}}^{\infty}dL\phi(L)L
\end{equation}
We adopt an $L_{\rm cut}$ value $\log L_{\rm cut}/{\rm erg~s^{-1}}=42.5$, which is around the characteristic luminosities at $z<6$. We present $\rho_{\mathrm{Ly}\alpha}$ value for each fitting function results at each redshift in Table \ref{table:lf_params_EoR}. \Add{We compare the redshift evolution of $\rho_{\mathrm{Ly}\alpha}$ to the UV luminosity density $\rho_{\rm UV}$. We adopt dust-uncorrected $\rho_{\rm UV}$ values integrated down to the absolute UV magnitude $M_{\rm UV}=-17.0$ mag from \cite{Bouwens22}. We show the redshift evolution of $\rho_{\mathrm{Ly}\alpha}$ and $\rho_{\rm UV}$ in Figure \ref{lum_den_evolution}. We confirm that $\rho_{\mathrm{Ly}\alpha}$ does not show strong redshift evolution at $z=2-6$ compared with that for $\rho_{\rm UV}$. We also confirm the $\rho_{\mathrm{Ly}\alpha}$ start decreasing from $z\sim5$ faster than $\rho_{\rm UV}$, suggesting the impact of cosmic reionization.} \Add{Particularly, $\rho_{\mathrm{Ly}\alpha}$ values drops by around one-third value between $z=7.0$ and 7.3. We fit the Ly$\alpha$ and UV luminosity density evolutions with a function proportional to $(1+z)^\xi$, where $\xi$ is a free parameter. We perform the fitting over the redshift ranges $z=5.7-7.0$ and $6.0-8.0$ for Ly$\alpha$ and UV luminosity densities, respectively. We show the fitted functions for Ly$\alpha$ and UV luminosity densities at the right panel of Figure \ref{lum_den_evolution}. We confirm steady decrease in the luminosity densities for Ly$\alpha$ and UV until at the redshift $z=7.0$ and 8.0, respectively. However, beyond those redshifts, the luminosity density evolution suddenly drops for both Ly$\alpha$ and UV. As noted in \cite{Konno14}, the redshift offset between the sudden drops in the luminosity densities of Ly$\alpha$ and UV suggests that the Ly$\alpha$ luminosity density drop between $z=7.0$ to 7.3 is not caused by the intrinsic galaxy evolution, but rather caused by the sudden IGM evolution at the corresponding redshift.}

\subsection{Characteristics of LAE Clustering}
\begin{figure}[htbp]
    \centering
    \includegraphics[width=0.9\linewidth]{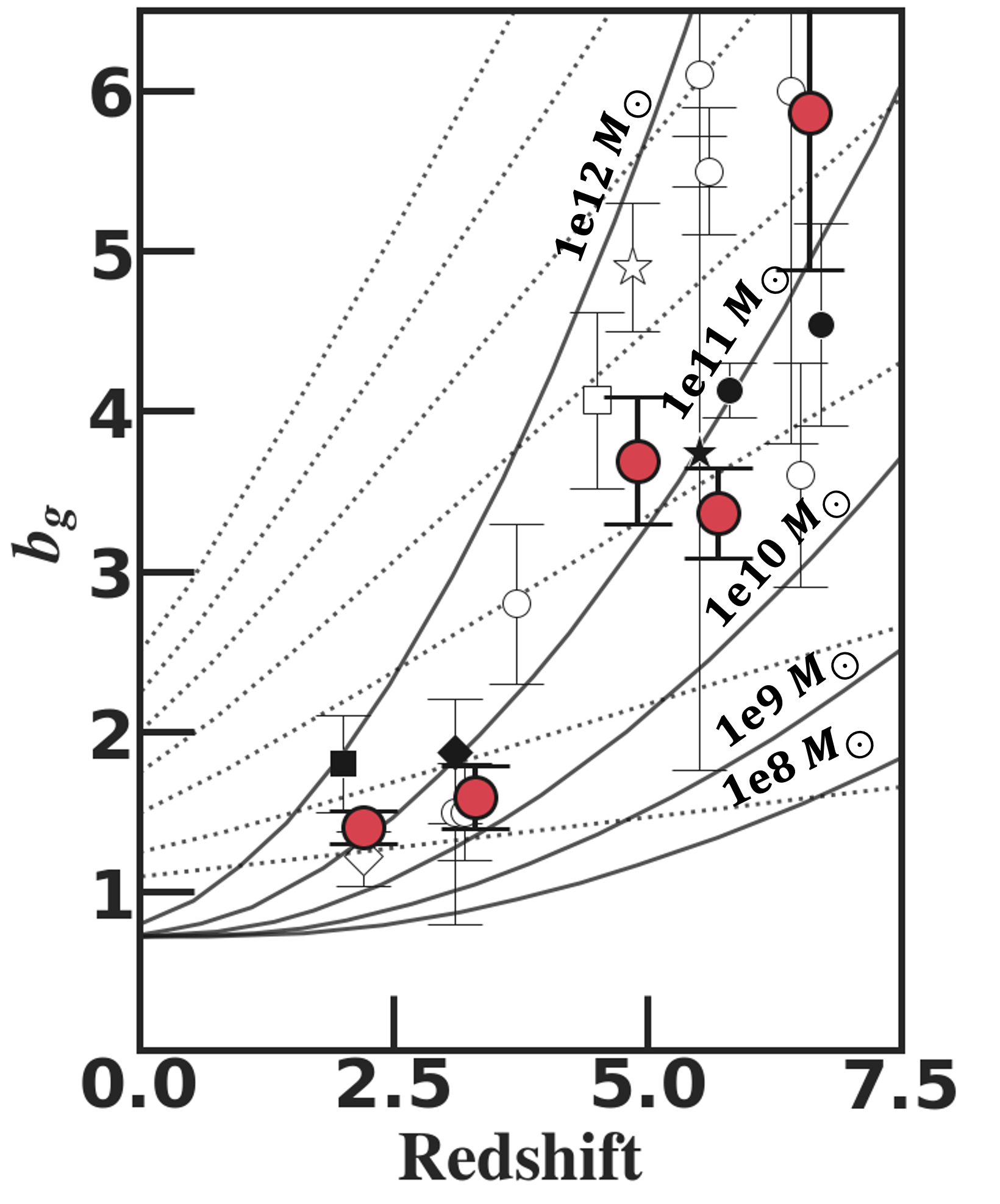}
    \caption{Redshift evolution of \Add{the bias of} LAEs as a function of redshift. The red circle represent average galaxy bias of our LAEs at $z=2.2$, 3.3, 4.9, 5.7, and 6.6. The black and white circles LAE bias value from \cite{Ouchi18} and \cite{Ouchi10}, respectively. The black and white stars represent LAE bias values obtained by \cite{Ouchi05} and \cite{Ouchi03}, respectively. The black and white squares are LAE bias values measured by \cite{Guaita10} and \cite{2007ApJ...668...15K}, respectively. The black and white diamonds represent LAE bias measured by \cite{2007ApJ...671..278G} and \Add{\cite{Kusakabe18}, respectively}. All the bias measurements are corrected to match $\sigma_8$ value consistent with \cite{Planck20}. The solid line indicate bias of the dark matter halos with a halo mass of $10^8$, $10^9$, $10^{10}$, $10^{11}$, and $10^{12}~M_{\odot}$ when each dark matter halo host one galaxy \citep{Ouchi18}. The dotted lines represent redshift evolution of bias in the case of galaxy-conserving model \citep{Fry96}.}
    \label{bias_evolution}
\end{figure}
To examine the redshift evolution of LAE clustering, we summarize the $b_{\rm ave}$ values at $z=2.2$ to $6.6$ in Figure \ref{bias_evolution}. We also \Add{present} LAE galaxy bias values from the literature including those from \cite{Ouchi18}, \cite{Ouchi10}, \cite{Ouchi05}, \cite{Ouchi03}, \cite{Guaita10}, \cite{2007ApJ...668...15K}, \cite{2007ApJ...671..278G}, and \Add{\cite{Kusakabe18}}. We do see a trend of the galaxy bias values increasing toward higher redshift as expected in the hierarchical structure formation from $\Lambda$CDM model. For the comparison, we plot the bias of the dark matter halos with a halo mass of $10^8$ to $10^{12}~M_\odot$ calculated in \cite{Ouchi18}. Our galaxy bias values at $z=2.2$ to 6.6 are roughly consistent with those expected for dark matter halos of mass $10^{11}~M_\odot$. We also overplot the bias evolution based on the galaxy-conserving model \citep{Fry96}. Galaxy-conserving model assumes the bias evolution based on the following formula:
\begin{equation}
b_{g}(z)=1 + (b_{g}(z=0) - 1)/D(z).
\label{gal_conserv}
\end{equation}
Here, $D(z)$ is the growth factor. According to the galaxy-conserving model, our LAEs at $z=4.9$ and $5.7$ lies around the evolutionary tracks with $b_{g}(z=0)\sim1.5$. As discussed in \cite{Ouchi18}, such a $b_{g}(z=0)$ value corresponds to that of $\sim6 L_\star$ massive bright galaxies in current universe \citep{Zehavi05}. \Add{We note that if the contamination fraction $f_{\rm cont}$ is non-negligible, the bias values systemically increase. For example, if we assume $f_{\rm cont}$ as high as it is assumed in \cite{Konno18} (i.e., 30\%) to be conservative (see Section \ref{lf:cont}), the bias values systematically increase by the factor of $1/(1-f_{\rm cont})\simeq1.5$ \citep[c.f., ][]{Ouchi18} and the inferred LAE halo masses increase correspondingly.}
\par

We \Add{would} like to highlight that we confirm a significant excess from the power-law model at $\theta<10$ arcsec at $z=2.2$\Add{, as seen in Figure \Add{\ref{acf}}}. We see tentative departure from the power-law function for the ACFs at $z=3.3$ and 5.7 as well. The similar features are seen in the ACFs for $z>0$ LBGs \citep[e.g.,][]{Ouchi05b}. The departures from the power-law function at small angular scales are often explained by the one-halo term, which is caused by the pairs of galaxies that reside in the same dark matter halo \citep{Hawkins03,Zehavi04,Ouchi05b,Harikane16}. The confirmation of the one halo term from ACFs of LAE further backs up the picture that LAE also form inside the dark matter halo just like LBGs. Another possible explanation for this excess is that the humps are caused by the multiple detections of a single extended Ly$\alpha$ blob (LAB). Because the sizes of LABs' extended Ly$\alpha$ emissions can reach $>100$ kpc \citep[e.g.,][]{Matsuda09}, corresponding to the angular separation of $>1$ arcsec at $z=2.2$, the multiple detection for the extended feature could mimic the one-halo term in the ACFs. \Add{To test whether LABs can explain the enhancement of clustering below 10 arcsec, we estimate the number count of LABs at $z=2.2$ based on the LAB number density at $z=2.3$ reported by \cite{Yang10}. We adopt the 1$\sigma$ upper limit of the LAB number density at $2.8\times10^{-5}~{\rm cMpc}^{-3}$ reported by \cite{Yang10}. If we assume each LAB contributes an LAE-LAE pair at below 10 arcsec, the contribution of LABs to the LAE-LAE pair counts below 10 arcsec scale is at most 20\% of the total pair counts for our $z=2.2$ LAE sample. Thus, other factors such as the one-halo term from non-LAB pairs are likely necessary to fully explain the humps seen in the $z=2.2$ LAE ACF at scales below 10 arcsec scale.} \par 
\subsection{Inferring the Ionization State of IGM}
We infer $x_{\rm \HI}$ from the redshift evolution of the luminosity function and angular auto-correlation functions. To have self-consistent $x_{\rm \HI}$ constraints between that from luminosity function and auto-correlation function, we compare these statistical properties with the predictions based on the same model. To predict the luminosity function and auto-correlation functions for different $x_{\rm \HI}$ in self-consistent manner, we use {\sc 21cmFAST v3} \citep[hereafter {\sc 21cmFAST};][]{Mesinger11,Murray20} to simulate the ionization of IGM and galaxy distributions and extract statistical properties of LAEs \citep[e.g., ][]{Sobacchi15,Mason18,Morales21}. We run {\sc 21cmFAST} in a similar manner to the EoS2016 simulations \citep{Mesinger16}, but with smaller box size of $(300)^3 {\rm cMpc}^3$ instead of ${1.6}^3 {\rm cGpc}^3$ for the computational efficiency. We set the density and ionization box resolutions ({\tt DIM} and {\tt HII\_DIM}, respectively) to {\tt DIM}=1024 and {\tt HII\_DIM}=258, respectively. To mock EoS2016 simulations, we set the ionizing efficiency $\zeta$ and minimum virial temperature for halos $T_{\rm vir, min}$ as $\zeta=20$ and $T_{\rm vir,min}=2\times10^4~{\rm K}$, respectively. We incorporated the effect from inhomogeneous recombination of ionized IGM by setting {\tt INHOMO\_RECO} true.\par
We then specify the galaxy UV luminosity to the generated dark matter halos based on the halo mass to UV luminosity relation based on \cite{Mason15}. We adopt the same assumption as the one for \cite{Mason18} to randomly generate the intrinsic Ly$\alpha$ line profile (e.g., velocity offset and dispersion of Ly$\alpha$, shape of the emission line) and Ly$\alpha$ EW distribution for different UV luminosity. We and \cite{Mason18} adopt the same intrinsic Ly$\alpha$ $EW_0$ distributions that are calibrated with the $z\sim6$ Ly$\alpha$ spectroscopic sample from \cite{Pentericci18} and \cite{DeBarros17}. \Add{We have confirmed the validity of this intrinsic $EW_0$ distribution assumption by finding the consistency between our $EW_{0,e}$ measurements at $z=4.9$ and 5.7 (i.e., $EW_{0,e}={38.5}_{-2.0}^{+2.2}$ and ${32.9}_{-0.7}^{+0.8}$ {\AA}, respectively) and the measurements from \cite{Pentericci18} (i.e., $EW_{0,e}={32}\pm{8}$ {\AA}).} \Add{We also generate the intrinsic velocity offset and the velocity dispersion of Ly$\alpha$ from the halo-mass dependent probability distribution based on the equation 4 of \cite{Mason18}}. We consider the IGM scattering of Ly$\alpha$ for individual galaxies based on the ionization state of the IGM on the sight-lines simulated by {\sc 21cmFAST}. \Add{We calculate the Ly$\alpha$ damping wing optical depth based on the formulation of \cite{ME98} for the given density and ionization state of the IGM on the line of sight}.\par
After considering IGM attenuation, we calculate Ly$\alpha$ luminosity function and auto-correlation function based on the simulation. We conduct {\sc 21cmFAST} simulations at global $x_{\rm \HI}$ values from 0 to 0.975 with step size of $\Delta x_{\rm \HI}=0.025$. 
\Add{Since the ACFs derived from narrow-band observations with a redshift coverage of $\Delta z = 0.1$ cannot distinguish between different morphological features in ionization maps \citep[e.g., as demonstrated in ][]{MF08a,McQuinn07}, we have opted to simply overlay ionization maps with different global neutral hydrogen fractions $x_{\rm HI}$ values from a single cosmic reionization scenario calculation onto the density field at redshifts $z = 4.9$, $5.7$, $6.6$, $7.0$, and $7.3$.} This approach allows us to calculate the statistical properties of LAEs. We calculate the Ly$\alpha$ luminosity function and ACFs in the randomly cutted out a sub-volume of the whole comoving simulation box with a radial length of $\Delta z=0.1$ to match the radial size of the actual survey volume by narrow-band filters. We conduct this procedure for 100 times and determine the mean and variance of the Ly$\alpha$ LFs and ACFs at each $x_{\rm \HI}$ and redshift.\par

\begin{figure*}[htbp]
    \includegraphics[width=\linewidth]{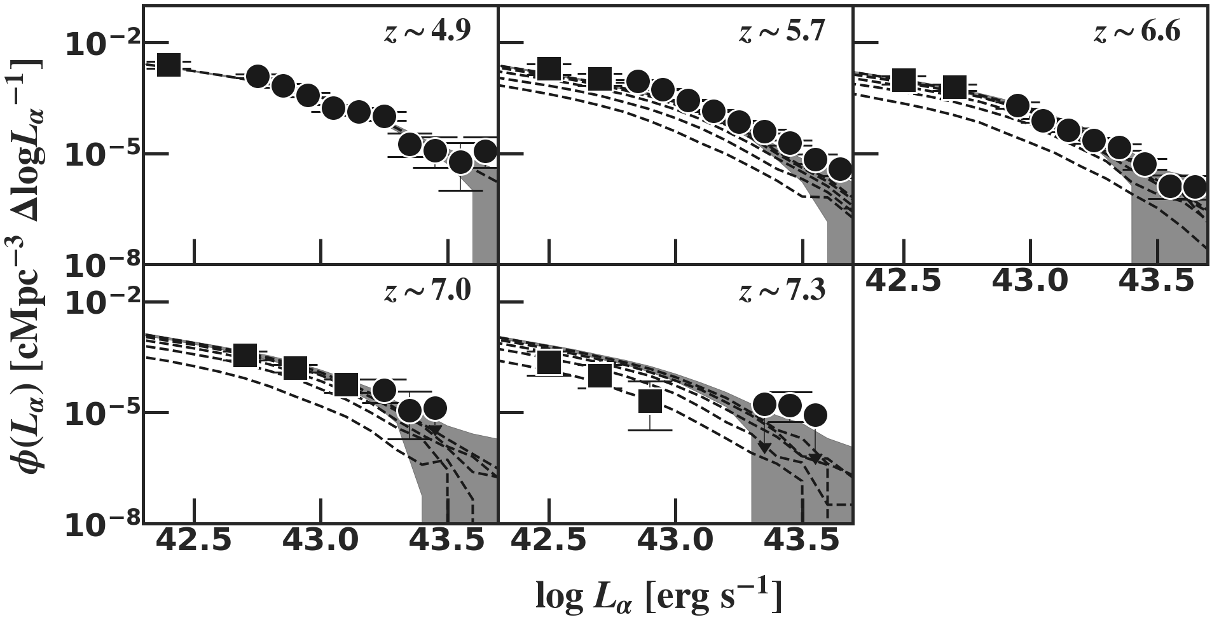}
    \caption{Comparison between observed Ly$\alpha$ luminosity functions and prediction by our {\sc 21cmFAST} simulations at $z=4.9, 5.7, 6.6, 7.0,$ and 7.3 for different $x_{\rm \HI}$. The black dots and the bars represents the observed luminosity function and their errors, respectively. \Add{On the upper left panel,} we show the predicted luminosity function and 68-th percentile uncertainty at \Add{$x_{\rm \HI}=0$} \Add{at $z=4.9$} in dotted line and grey shade, respectively. For the other redshifts, we show the luminosity function at $x_{\rm \HI}=0.0, 0.2, 0.4, 0.6,$ and 0.8. The luminosity function monotonically decreases with $x_{\rm \HI}$ increases. We show 68-th percentile uncertainty for luminosity functions at $x_{\rm \HI}=0.0$ as references.}
    \label{lf_xHI}
\end{figure*}

When we calculate Ly$\alpha$ luminosity function from the simulation, we adopt similar methodology as adopted in \cite{Morales21}. We assume UV luminosity function in \cite{Harikane22} and convert the UV luminosity function into Ly$\alpha$ luminosity function by randomly assigning the Ly$\alpha$ luminosity based on Ly$\alpha$ equivalent width distribution by UV magnitudes. For the luminosity function, we only consider the luminosity range of $\log L_\alpha\Add{/{\rm erg~s^{-1}}}=$42.4$-$43.4. The lower limit of the luminosity range is set to ensure that the resolution of the simulation does not affect the prediction, whereas the upper limit is set to avoid the possible contamination by AGNs at $\log L_\alpha>43.5$ erg/s \citep[e.g.,][]{Konno16}. We show the comparison between the simulated prediction of Ly$\alpha$ luminosity function for different $x_{\rm \HI}$ at $z=4.9$, 5.7, 6.6, 7.0, and 7.3 in the Figure \ref{lf_xHI}. In the upper left panel of Figure \ref{lf_xHI}, we present observed Ly$\alpha$ LF at $z=4.9$ together with model prediction for Ly$\alpha$ LF at \Add{$x_{\rm \HI}=0$} to check that the intrinsic Ly$\alpha$ LF assumed in model is consistent with the observed LFs after the completion of cosmic reionization. We confirm that the observed LFs and model predictions are consistent within errors. In other panels of Figure \ref{lf_xHI}, we compare the observed and predicted model Ly$\alpha$ LFs for $x_{\rm \HI}=0.0,$ 0.2, 0.4, 0.6, and 0.8. The Ly$\alpha$ LF monotonically decreases as $x_{\rm \HI}$ \Add{increases} for the luminosity range $\log L_\alpha\Add{/{\rm erg~s^{-1}}}=42.4-43.4$ at each redshift due to IGM attenuation. We calculate posterior probability density at different $x_{\rm \HI}$ values by assuming uniform prior for $x_{\rm \HI}$ and likelihood function based in the gaussian form. Based on the calculated posterior probability distribution, we infer the 97.5-th percentile upper limit of $x_{\rm \HI}<0.05$ at $z=5.7$. For $z=6.6$, 7.0, and 7.3, we adopt median (16-/84-th) percentile values as the best-fit (uncertainty) values. We find \Add{$x_{\rm \HI}=0.15^{+0.10}_{-0.08}$, $0.18^{+0.14}_{-0.12}$, and $0.75^{+0.09}_{-0.13}$} at $z=6.6$, 7.0, and 7.3, respectively.\par

\begin{figure*}[htbp]
    \includegraphics[width=\linewidth]{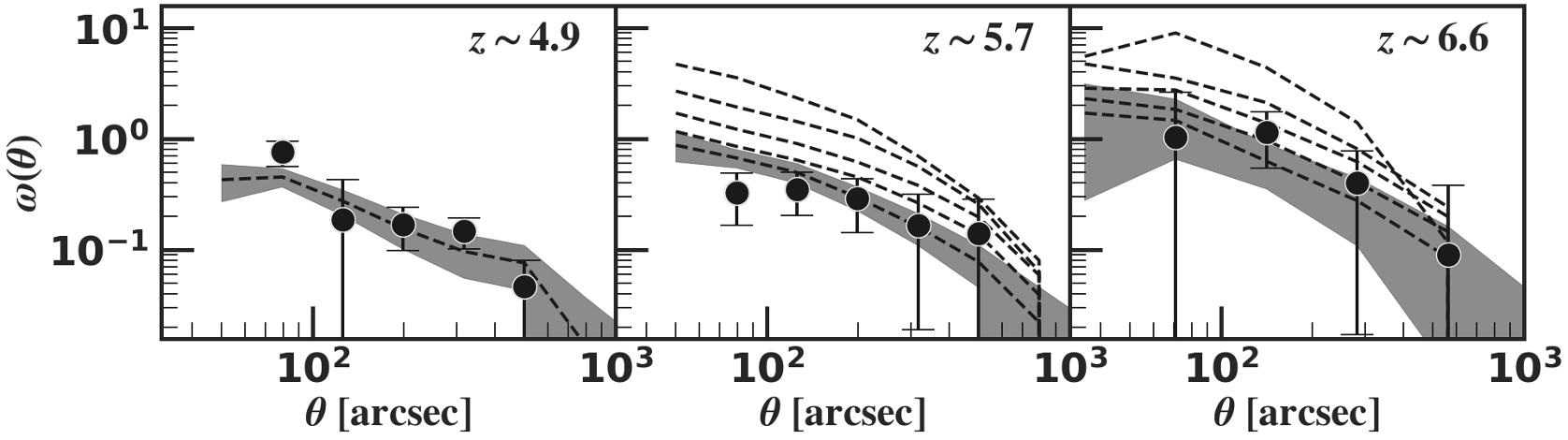}
    \caption{Comparison between observed Ly$\alpha$ emitter's two-point angular correlation functions and prediction by our {\sc 21cmFAST} simulations at $z=4.9, 5.7$ and 6.6 for different $x_{\rm \HI}$. The black (white) dots and the bars represents the observed angular correlation function and their errors (not) used for $x_{\rm \HI}$ inference, respectively. \Add{On the upper left panel,} we show the predicted \Add{angular correlation function} and 68-th percentile uncertainty at $x_{\rm \HI}=0$ at $z=4.9$ in dotted line and grey shade, respectively. For the other redshifts, we show the \Add{angular correlation function} at $x_{\rm \HI}=0.0, 0.2, 0.4, 0.6,$ and 0.8. The angular correlation function monotonically increases with $x_{\rm \HI}$ increase.}
    \label{acf_xHI}
\end{figure*}

For the ACF, we consider LAE with the threshold Ly$\alpha$ luminosity ($\log L_{\alpha, \rm th}$) of $\log L_{\alpha, \rm th}/{\rm erg~s^{-1}}=42.6,42.7,$ and 42.8 assuming $EW_0\approx20$ {\AA} at $z=4.9, 5.7$, and 6.6, respectively, to match the \Add{luminosity cut we applied for the observed ACFs in Section \ref{acf_expl}}. When we infer $x_{\rm \HI}$ from the ACFs, we only consider $1.7\leq\log \theta/{\rm arcsec}\leq2.8$. For the model, we do not consider IC correction as IC values for observed ACFs at $z=4.9$ to $6.6$ (i.e., $\sim2\times10^{-2}$) are negligible compared with $\omega(\theta)$ values at $1.5\leq\log \theta/{\rm arcsec}\leq2.3$. In the left panel of Figure \ref{acf_xHI}, we present observed ACFs at $z=4.9$ together with model prediction for the LAE ACF at \Add{$x_{\rm \HI}=0.0$} to check that the intrinsic ACFs assumed in model is consistent with the observed LFs after the completion of cosmic reionization. We confirm that our simulated prediction are consistent with the observed ACFs within model's uncertainties. On the other panels of Figure \ref{acf_xHI}, we show our ACF predictions at $x_{\rm \HI}=0.0$, 0.2, 0.4, 0.6, and 0.8 together with observed ACFs. The simulated ACFs values at $1.5\leq\log \theta/{\rm arcsec}\leq2.3$ monotonically increases as $x_{\rm \HI}$ increases. We see that the ACF at $z=5.7$ are also consistent with prediction at \Add{$x_{\rm \HI}=0.0$}, while ACF at $z=6.6$ departs from the prediction at $x_{\rm \HI}=0$, suggesting the impact of non-negligible IGM attenuation on observed ACFs. \Add{We have also compared our simulated ACFs at $z=6.6$ using {\sc 21cmFAST} to the prediction based on $N$-body radiative transfer simulation from \cite{McQuinn07}. We confirm that the ACF predictions from \cite{McQuinn07} assuming similar LAE halo mass (i.e., $>7\times10^{10}~M_{\odot}$) as our LAE sample with different $x_{\rm \HI}$ match within errorbars with the prediction in this work.} Similar to the case of Ly$\alpha$ LFs, we calculate posterior probability density at \Add{different} $x_{\rm \HI}$ by assuming uniform prior on $x_{\rm \HI}$ and likelihood function based in the gaussian form. For ACFs, we consider non-diagonal component of the covariance matrix for both model and observation. Based on the calculated posterior probability distribution, we infer the $x_{\rm \HI}$ by adopting median (16-/84-th) percentile values of the posterior probability density distribution as the best-fit (uncertainty) values. We find \Add{$x_{\rm \HI}={0.06}^{+0.12}_{-0.03}$ and ${0.21}^{+0.19}_{-0.14}$} at $z=5.7$ and 6.6, respectively. We find that inferred $x_{\rm \HI}$ values from ACF are slightly higher those from LFs, but they remain consistent within errorbars.\par

\begin{figure*}[htbp]
\centering
\begin{center}
\includegraphics[width=\linewidth]{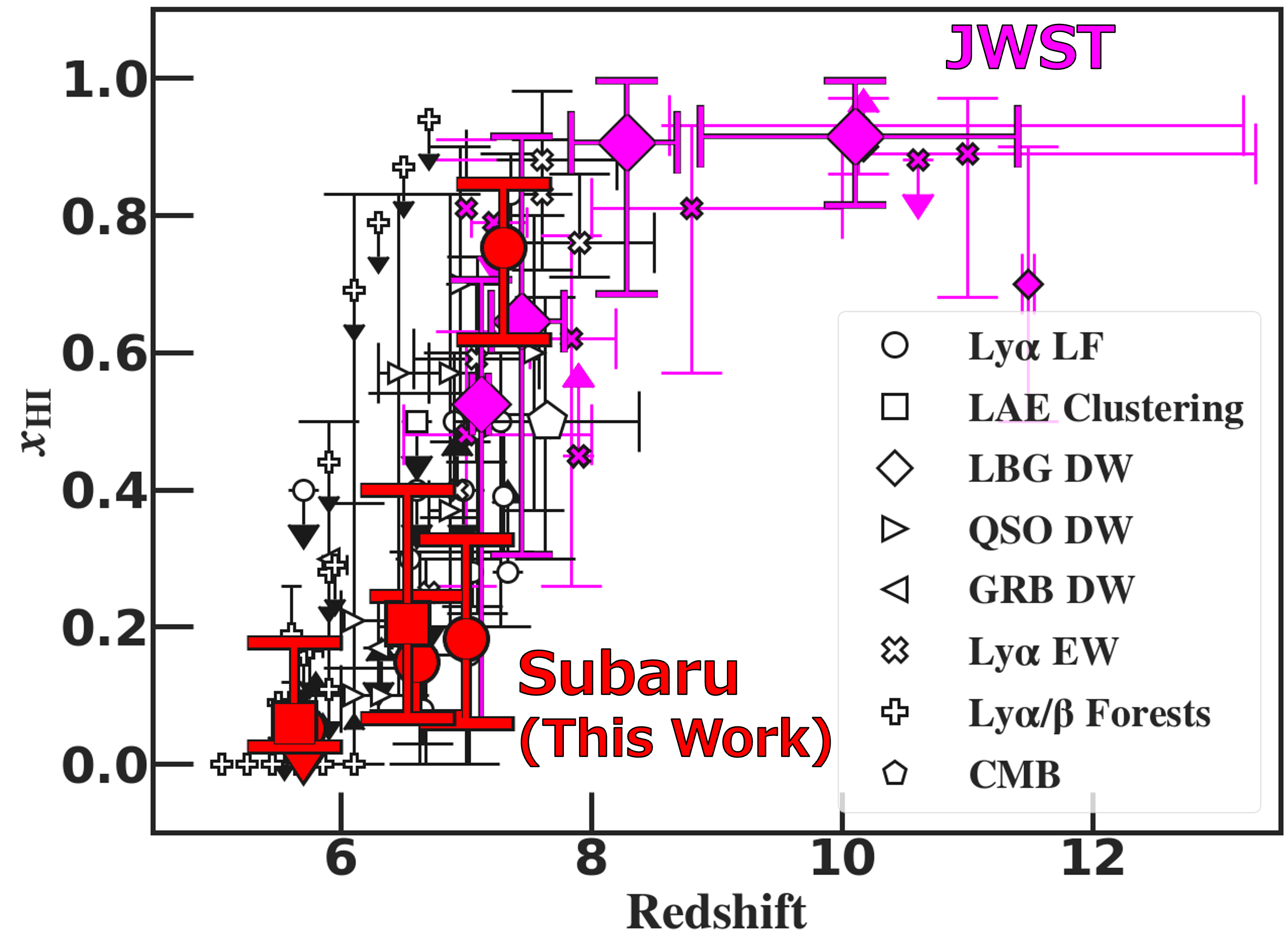}
\end{center}
\caption{\Add{Redshift evolution of $x_{\rm \HI}$. The red circles and bars represent our $x_{\rm \HI}$ estimates and uncertainties based on the Ly$\alpha$ luminosity functions and LAE clusterings, respectively.}
Beside $x_{\rm \HI}$ estimates from this work, we also present $x_{\rm \HI}$ estimate from literature using Ly$\alpha$ luminosity function \citep[circles;][]{Ouchi10,Konno14,Zheng17,2018PASJ...70...55I,Morales21,Goto21,Ning22}, LAE clustering \citep[squares][]{Sobacchi15,Ouchi18}, Ly$\alpha$ damping wing measurement of LBGs \cite[diamonds;][]{Umeda24,CL23,Hsiao23}, damping wing measurements of QSOs \cite[right-tipped triangles;][]{2013MNRAS.428.3058S,2018ApJ...864..142D,2019MNRAS.484.5094G,2020ApJ...896...23W,Durovcikova23}, damping wing measurements of GRBs \citep[left-tipped triangles;][]{2006PASJ...58..485T,2014PASJ...66...63T}, Ly$\alpha$ equivalent width distributions \citep[X marks;][]{2015MNRAS.446..566M,2019ApJ...878...12H,2019MNRAS.485.3947M,2020ApJ...904..144J,2020MNRAS.495.3602W,Bolan22,Br23,Mo23,Nakane24,Tang24,Jones24}, Ly$\alpha$ forests and/or Ly$\alpha$+$\beta$ dark fraction/gaps measurements \citep[pluses;][]{2006AJ....132..117F,2015MNRAS.446..566M,2023ApJ...942...59J,2022ApJ...932...76Z,Zhu24,Spina24}, and the electron scattering of CMB \citep[pentagon;][]{Planck20}. \Add{The pink symbols represent the estimates based on JWST data \citep{Umeda24,CL23,Hsiao23,Mo23,Nakane24,Tang24,Jones24}.}}
\label{history}
\end{figure*}

\begin{figure}[htbp]
\centering
\begin{center}
\includegraphics[width=\linewidth]{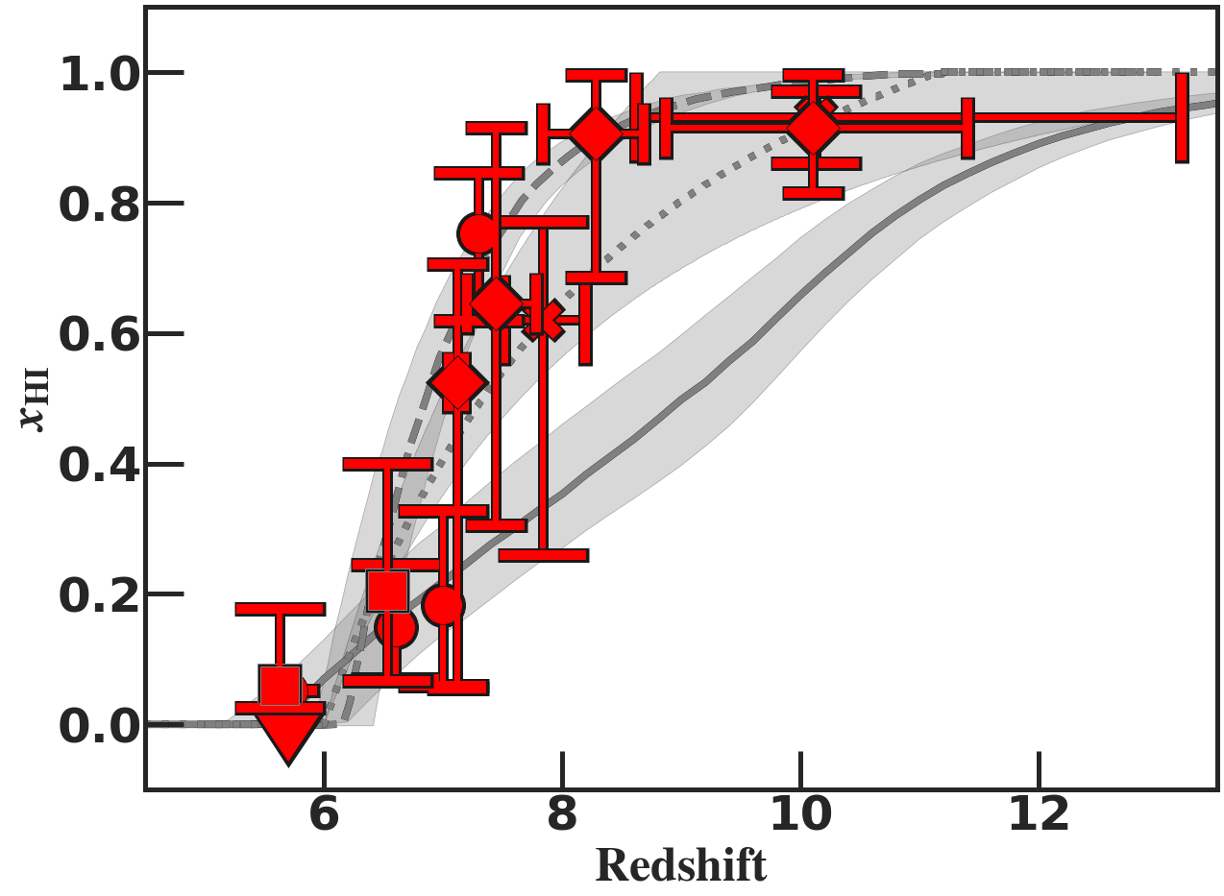}
\end{center}
\caption{\Add{Comparison between cosmic reionization scenarios and $x_{\rm \HI}$ estimates. The solid, dotted, and dashed lines and shades represent $x_{\rm \HI}$ evolution prediction from \cite{F19}, \cite{I18}, and \cite{N20}, respectively. The red circles and squares represent $x_{\rm \HI}$ from this work based on Ly$\alpha$ luminosity functions and LAE clusterings, respectively. The red diamonds and x-marks represent the $x_{\rm \HI}$ estimates from LBG damping wing by \cite{Umeda24} and Ly$\alpha$ equivalent width distributions by \cite{Nakane24}, respectively.}}
\label{history_compare}
\end{figure}

We summarize our $x_{\rm \HI}$ constraints in Figure \ref{history}. \Add{We also present $x_{\rm \HI}$ values from literature.} We present inferred $x_{\rm \HI}$ from Ly$\alpha$ LFs and clusterings in red circles and squares, respectively. As discussed in Section \ref{z_lf}, we find that the $x_{\rm \HI}$ remains low ($x_{\rm \HI}\lesssim0.2$) at the $z=5-7$, implying small evolution of IGM attenuation. However, \Add{our inferred $x_{\rm \HI}$ values} suddenly increase at $z=7.3$ up to $x_{\rm \HI}\simeq0.8$, \Add{consistent with the rapid decline of $\rho_{\rm Ly\alpha}$ values from $z=7.0$ to 7.3}. Recent $x_{\rm \HI}$ estimates from JWST suggests that the universe is almost completely neutral $x_{\rm \HI}\sim0.9$ at $z\simeq8-9$ \citep[e.g., ][]{Umeda24,Nakane24,Tang24}. By connecting the $x_{\rm \HI}$ values at $z=5.7-7.3$ inferred in this work \Add{based on LAEs} with those from JWST observations, we can draw the cosmic reionization picture that the major epoch of reionization occurred around $z\sim7-8$. \Add{In Figure \ref{history_compare}, we compare multiple cosmic reionization scenarios with $x_{\rm \HI}$ estimates from this work and representative values from literature based on JWST data. We show predictions from \cite{N20}, \cite{I18}, and \cite{F19} as representatives for the rapid, moderate, and extended cosmic reionization scenarios driven by the massive, all, and faint galaxies, respectively.} \cite{N20} suggest such a rapid cosmic reionization driven mostly by massive galaxies with stellar mass $\log M_\star/M_\odot>8$, reaching $x_{\rm \HI}=0.5$ at $z<7$ and $x_{\rm \HI}\sim0$ at $z\sim6$. However, redshift evolution of $x_{\rm \HI}$ inferred from this work and JWST results suggest that cosmic reionization occurs at \Add{slightly} earlier than predicted by \cite{N20}, \Add{while our $x_{\rm \HI}$ measurements do not have tension with the CMB electron scattering measurement as that for the extremely early reionization scenario claimed by \cite{Munoz24}.} The recent discovery of the overabundant UV-bright galaxies at $z>7$ suggested by JWST observations \Add{could hints at the elevated contributions from massive galaxies on ionizing photon budget at the early stage of EoR.} \citep[e.g.,][]{Harikane24}. Moreover, the lack of strong emission line detections for those $z>10$ could indicate a high ionizing photon escape fraction for these massive galaxies \citep[e.g.,][]{CL23,Carniani24}, enabling earlier progress of cosmic reionization.\par
\Add{Further investigations on the ionizing photon production efficiencies and escape fractions of high-redshift galaxies are crucial to confirm the rapid cosmic reionization scenario suggested in this work \citep[cf.][]{Simmonds24}.} \Add{Another interest is to reconcile the cosmic reionization history consistently with the Ly$\alpha$ forest measurements at $z<6$. \cite{Bosman24} use the Ly$\alpha$ forest measurements to constrain the UV background at $z<6$ to infer galaxy ionizing properties. \cite{Bosman24} infer the product between ionizing photon production efficiency and escape fraction (i.e., $\langle \xi_{\rm ion}f_{\rm esc}\rangle$) and find no significant discrepancies between the prediction assuming the bright LAE (i.e., $M_{\rm UV}<-17$) dominating the ionizing photon sources \citep{Matthee24} resembling the Naidu et al.'s cosmic reionization scenario. In contrast, $x_{\rm \HI}$ evolutions between $z=7.0-7.3$ from this work is more rapid than that of the posterior inference on cosmic reionization scenarios using Ly$\alpha$ forest, CMB, and UV luminosity function measurements \citep[Y. Qin et al. in preparation; see][]{Zhu24}. Explaining observables from Ly$\alpha$ forest and high-z LAEs self-consistently needs to be further investigated via theories and observations.}\par

\begin{deluxetable}{ccc}
\tablecolumns{3}
\tabletypesize{\scriptsize}
\tablecaption{Inferred $x_{\rm \HI}$ at different Redshifts
\label{table:xHI}}
\tablehead{
\colhead{$z$} & \colhead{Inference Method} & \colhead{$x_{\rm \HI}$}  \\
(1) & (2) & (3)
}
\startdata 
5.7 & LF & \Add{$<{0.05}$} \\
5.7 & ACF & \Add{${0.06}^{+0.12}_{-0.03}$} \\\hline
6.6 & LF & \Add{$0.15^{+0.10}_{-0.08}$} \\
6.6 & ACF & \Add{${0.21}^{+0.19}_{-0.14}$} \\\hline
7.0 & LF & \Add{$0.18^{+0.14}_{-0.12}$} \\\hline
7.3 & LF & \Add{$0.75^{+0.09}_{-0.13}$} \\
\enddata
\tablecomments{(1): Redshifts (2): Method used for ${x_{\rm \HI}}$ inference. LF and ACF represents the inference based on the luminosity function and the auto-correlation function, respectively. (3): Inferred $x_{\rm \HI}$ values at each redshift. The 97.5-th percentile value is given for $x_{\rm \HI}$ constraint by LF at $z=5.7$. The other $x_{\rm \HI}$ values correspond to the median and 16/84-th percentile values.}
\end{deluxetable}


\section{Conclusions} \label{sec:conc}
In this work, we present the Ly$\alpha$ luminosity function and the clustering properties of LAEs at $z=2.2-7.3$. We use the recent LAE sample constructed based on the deep optical imaging data from HSC-SSP and CHORUS surveys to calculate the statistical properties of LAEs at different redshift. We summarize our findings as follows:
\begin{enumerate}

\item Based on our LAE samples, we determine the Ly$\alpha$ LFs at $z=2.2$, 3.3, 4.9, 5.7, 6.6, 7.0, and 7.3. For $z\leq4.9$, we determine the best-fit Schechter/DPL parameters for the galaxy component and DPL parameters for the AGN component of the Ly$\alpha$ LFs. For $z\geq4.9$, we determined the best values for the Schechter/DPL functions. We reconfirm that there is significant evolution of the Schechter parameters at $z=7.3$, suggesting the accelerated evolution due to cosmic reionization.

\item We calculate the two-point angular auto-correlation function at $z=2.2$, 3.3, 4.9, 5.7, and 6.6. Based on the average galaxy bias values, the dark matter halos hosting LAEs are likely to evolve into the dark matter halo with masses of $\sim10^{11}~M_\odot$ at $z=0$, corresponding to super-$L^\star$ galaxies in the current universe.

\item We compare the Ly$\alpha$ LFs and ACF at $z\geq4.9$ with the cosmic reionization simulation predictions that considers the inhomogeneity of IGM. By comparing the model predictions at different $x_{\rm \HI}$, we infer $x_{\rm \HI}$\Add{$<0.05$ ($={0.06}^{+0.12}_{-0.03}$), $0.15^{+0.10}_{-0.08}$ (${0.21}^{+0.19}_{-0.14}$), $0.18^{+0.14}_{-0.12}$, and $0.75^{+0.09}_{-0.13}$} at $z=5.7$, $6.6$, $7.0$, and $7.3$ from Ly$\alpha$ LFs (ACFs), respectively.

\item We summarize the inferred $x_{\rm \HI}$ values at $z\geq5.7$ together with other $x_{\rm \HI}$ measurements including those inferred using recent JWST galaxy observations. We find that our $x_{\rm \HI}$ estimates \Add{and model-independent $\rho_{Ly\alpha}$ values} evolve rapidly around $z\sim7-8$, indicating that the majority of the epoch of reionization occurs around that redshift. This early and rapid progress in cosmic reionization could be due to the emergence of significant ionizing photon sources (i.e., UV-bright massive galaxies discovered by JWST) in the early universe.
\end{enumerate}

\section*{Acknowledgements}
We thank Brant Robertson, Rieko Momose, Yechi Zhang, Yuxiang Qin, Charlotte Mason, Sarah Bosman, Frederick Davies, Martin Haehnelt, Sandro Tacchella, Matthew Hayes, Shintaro Yoshiura, Hideyuki Yajima, Kenji Hasegawa, Kentaro Nagamine, Makoto Ando, Byeongha Moon, Thi Tran Thai, Chenghao Zhu, and participants of Kochel CLAW for useful discussions on this work. \par
The Hyper Suprime-Cam (HSC) collaboration includes
the astronomical communities of Japan and Taiwan,
and Princeton University. The HSC instrumentation
and software were developed by the National Astronomical
Observatory of Japan (NAOJ), the Kavli
Institute for the Physics and Mathematics of the Universe
(Kavli IPMU), the University of Tokyo, the High
Energy Accelerator Research Organization (KEK), the
Academia Sinica Institute for Astronomy and Astrophysics
in Taiwan (ASIAA), and Princeton University. 
Funding was contributed by the FIRST program from
the Japanese Cabinet Office, the Ministry of Education,
Culture, Sports, Science and Technology (MEXT),
the Japan Society for the Promotion of Science (JSPS),
Japan Science and Technology Agency (JST), the Toray
Science Foundation, NAOJ, Kavli IPMU, KEK, ASIAA,
and Princeton University. \par
This work is based on data collected at the Subaru
Telescope and retrieved from the HSC data archive system,
which is operated by Subaru Telescope and Astronomy
Data Center (ADC) at NAOJ. Data analysis was
in part carried out with the cooperation of Center for
Computational Astrophysics (CfCA) at NAOJ. We are
honored and grateful for the opportunity of observing
the Universe from Maunakea, which has the cultural,
historical and natural significance in Hawaii.\par
This work is supported by World Premier International Research Center Initiative (WPI Initative), MEXT, Japan, KAKENHI (20H00180, 21H04467, 23KJ0646, 24K17084, 24H00245, 23H00131, 24KJ0058, 24K17101). U.H., A.M., and Y.K. are supported by The Forefront Physics and Mathematics Program to Drive Transformation (FoPM). U.H. and A.M. are supported by the JSPS Research Fellowship for Young Scientists. \Add{S.S. acknowledges support for this work from NSF-2219212 and the U.S. Department of Energy, Office of Science, Office of High Energy Physics under DE-SC0024694. K.S. is supported by the Toray Science Foundation.}
The English writing in this paper was improved with ChatGPT (OpenAI 202026), while no sentences were generated \Add{by ChatGPT} from scratch.\par
{\it Facitilies}: Subaru (HSC)

\bibliography{manuscript.bib}{}
\bibliographystyle{aasjournal}

\end{document}